\documentstyle[12pt]{article}%
\newlength\Cscr\newlength\Csave
\newlength\Ctenthex\newlength\CFxsize
\newlength\CFxsizeps\newlength\CFsizemakebox\newlength\CFleftcrop
\newlength\CFrightcrop\newlength\CZtbldist\newlength\CZfigdist
\newlength
\CGDnum\newlength\CGDtext\newcounter{Cscr}%
\newcounter{CBcit}\newcounter{CElett}%
\newcounter{CBauthornu%
m}\newcounter{Ceqindent}\newcounter{CBtnc}
\newcounter{CBtntc}\newcounter{CEht}%
\newcounter{Cbscurr}\newcounter{CbsA}\newcounter{CbsB}\newcounter
{CbsC}\newcounter{CbsD}\hoffset\Cscr\voffset
\Cscr\protect
\begin{document}\setcounter{figure}{1}\setcounter{t%
able}{0}\renewcommand\theequation{\arabic{equation}}\renewcommand
\thetable{\arabic{table}}\renewcommand\thefigure{\arabic{figure}%
}\renewcommand\thesection{\arabic{section}}\renewcommand
\thesubsection{\arabic{section}.\arabic{subsection}}\renewcommand
\thesubsubsection{\arabic{section}.\arabic{subsection}.\arabic{s%
ubsubsection}}\setcounter{CEht}{10}\setcounter{CbsA}{1}%
\setcounter{CbsB}{1}\setcounter{CbsC}{1}\setcounter{CbsD}{1}{%
\centering{\protect\mbox{}}\\*[\baselineskip]{\large\bf A new pr%
oposal for the fermion doubling problem\\[0.4ex]II.~Improving th%
e operators for finite lattices}\\*}{\centering{\protect\mbox{}}%
\\John P.~Costella\vspace{1ex}\\*}{\centering{\small\mbox{}%
\protect\/{\protect\em Faculty of Mathematics, Mentone Grammar, 
63 Venice Street, Mentone, Victoria 3194, Australia\protect\/}}%
\\}{\centering{\small\mbox{}\protect\/{\protect\em jpcostella@ho%
tmail.com;\hspace{1ex} jpc@mentonegs.vic.edu.au;\hspace{1ex} jpc%
@physics.unimelb.edu.au\protect\/}}\\}{\centering{\protect\mbox{%
}}\\(24\ July 2002\vspace{1ex})\\}\par\vspace\baselineskip
\setlength{\Csave}{\parskip}{\centering\small\bf Abstract\\}%
\setlength{\parskip}{-\baselineskip}\begin{quote}\setlength{%
\parskip}{0.5\baselineskip}\small\noindent In a previous paper I 
showed how the ideal SLAC derivative and second-derivative opera%
tors for an infinite lattice can be obtained in simple closed fo%
rm in position space, and implemented very efficiently in a stoc%
hastic fashion for practical calculations on finite lattices. In 
this second paper I show how the small (order $1/N$) errors intr%
oduced by truncating the operators to a finite lattice may be re%
moved by a small adjustment of coefficients, without incurring a%
ny additional computational cost. The derivation of these result%
s is again presented in a simple, pedagogical fashion. \end{quot%
e}\setlength{\parskip}{\Csave}\par\refstepcounter{section}\vspace
{1.5\baselineskip}\par{\centering\bf\thesection. Truncating the 
SLAC derivative operators to fit on a finite lattice\\*[0.5%
\baselineskip]}\protect\indent\label{sect:Truncation}In a previo%
us paper~[\ref{cit:Costella2002}] we looked at the ideal ``SLAC%
'' specification of \protect\ref{au:Drell1976}~[\ref{cit:Drell19%
76}] for the spatial derivative operator on an infinite one-dime%
nsional lattice, which demands that $-i$ times the Fourier trans%
form of the derivative operator take on its ideal functional for%
m, \setcounter{Ceqindent}{0}\protect\begin{eqnarray}\hspace{-1.3%
ex}&\displaystyle p_{\mbox{\scriptsize$\:\!$ideal}}=p,\protect
\nonumber\setlength{\Cscr}{\value{CEht}\Ctenthex}\addtolength{%
\Cscr}{-1.0ex}\protect\raisebox{0ex}[\value{CEht}\Ctenthex][\Cscr
]{}\protect\end{eqnarray}\setcounter{CEht}{10}within the first B%
rillouin zone. Such an operator avoids the pathologies of the fe%
rmion doubling problem or the violation of chiral invariance tha%
t have plagued other sub-optimal definitions of the derivative o%
perator, but has a representation in position space that is very 
``nonlocal''\mbox{$\!$}, \setcounter{Ceqindent}{0}\protect\begin
{eqnarray}\protect\left.\protect\begin{array}{rcl}\protect
\displaystyle{\protect\it\Delta\!\:}_{\mbox{\scriptsize ideal}}f%
(x)\hspace{-1.3ex}&\protect\displaystyle=&\hspace{-1.3ex}\protect
\displaystyle\mbox{$\protect\displaystyle\protect\frac{1}{a}$}%
\setcounter{Cbscurr}{25}\setlength{\Cscr}{\value{Cbscurr}%
\Ctenthex}\addtolength{\Cscr}{-1.0ex}\protect\raisebox{0ex}[%
\value{Cbscurr}\Ctenthex][\Cscr]{}\hspace{-0ex}{\protect\left\{%
\setlength{\Cscr}{\value{Cbscurr}\Ctenthex}\addtolength{\Cscr}{-%
1.0ex}\protect\raisebox{0ex}[\value{Cbscurr}\Ctenthex][\Cscr]{}%
\protect\right.}\hspace{-0.25ex}\setlength{\Cscr}{\value{Cbscurr%
}\Ctenthex}\addtolength{\Cscr}{-1.0ex}\protect\raisebox{0ex}[%
\value{Cbscurr}\Ctenthex][\Cscr]{}\setcounter{CbsD}{\value{CbsC}%
}\setcounter{CbsC}{\value{CbsB}}\setcounter{CbsB}{\value{CbsA}}%
\setcounter{CbsA}{\value{Cbscurr}}\ldots-\mbox{$\protect
\displaystyle\protect\frac{1}{4}$}f(x+4a)+\mbox{$\protect
\displaystyle\protect\frac{1}{3}$}f(x+3a)-\mbox{$\protect
\displaystyle\protect\frac{1}{2}$}f(x+2a)\setcounter{Ceqindent}{%
200}\setlength{\Cscr}{\value{CEht}\Ctenthex}\addtolength{\Cscr}{%
-1.0ex}\protect\raisebox{0ex}[\value{CEht}\Ctenthex][\Cscr]{}\\*%
[0.55ex]\protect\displaystyle\hspace{-1.3ex}&\protect
\displaystyle&\hspace{-1.3ex}\protect\displaystyle{\protect\mbox
{}}\hspace{\value{Ceqindent}\Ctenthex}\setcounter{CEht}{30}+f(x+%
a)-f(x-a)\setcounter{Ceqindent}{150}\setlength{\Cscr}{\value{CEh%
t}\Ctenthex}\addtolength{\Cscr}{-1.0ex}\protect\raisebox{0ex}[%
\value{CEht}\Ctenthex][\Cscr]{}\\*[0.55ex]\protect\displaystyle
\hspace{-1.3ex}&\protect\displaystyle&\hspace{-1.3ex}\protect
\displaystyle{\protect\mbox{}}\hspace{\value{Ceqindent}\Ctenthex
}+\mbox{$\protect\displaystyle\protect\frac{1}{2}$}f(x-2a)-\mbox
{$\protect\displaystyle\protect\frac{1}{3}$}f(x-3a)+\mbox{$%
\protect\displaystyle\protect\frac{1}{4}$}f(x-4a)-\ldots
\setlength{\Cscr}{\value{CbsA}\Ctenthex}\addtolength{\Cscr}{-1.0%
ex}\protect\raisebox{0ex}[\value{CbsA}\Ctenthex][\Cscr]{}\hspace
{-0.25ex}{\protect\left.\setlength{\Cscr}{\value{CbsA}\Ctenthex}%
\addtolength{\Cscr}{-1.0ex}\protect\raisebox{0ex}[\value{CbsA}%
\Ctenthex][\Cscr]{}\protect\right\}}\hspace{-0ex}\setlength{\Cscr
}{\value{CbsA}\Ctenthex}\addtolength{\Cscr}{-1.0ex}\protect
\raisebox{0ex}[\value{CbsA}\Ctenthex][\Cscr]{}\setcounter{CbsA}{%
\value{CbsB}}\setcounter{CbsB}{\value{CbsC}}\setcounter{CbsC}{%
\value{CbsD}}\setcounter{CbsD}{1},\setlength{\Cscr}{\value{CEht}%
\Ctenthex}\addtolength{\Cscr}{-1.0ex}\protect\raisebox{0ex}[%
\value{CEht}\Ctenthex][\Cscr]{}\protect\end{array}\protect\right
.\protect\label{eq:Truncation-DeIdeal}\protect\end{eqnarray}%
\setcounter{CEht}{10}and indeed links the lattice site at which 
we wish to take the derivative to every other site in the (one-d%
imensional) lattice. This has, historically, presented a barrier 
to the practical implementation of the SLAC operator in any larg%
e-scale lattice calculations.\par In [\ref{cit:Costella2002}]\ I 
proposed a solution to this computational barrier, namely, that 
the SLAC derivative operator can be implemented in a \mbox{}%
\protect\/{\protect\em stochastic\protect\/} fashion, so that fo%
r a one-dimensional finite lattice of $N$ sites the average numb%
er of computations required to implement each derivative operati%
on is of order $\log N$ rather than of order~$N$.\par However, e%
ven though the use of the ideal SLAC specification ensures that 
the representation (\protect\ref{eq:Truncation-DeIdeal}) of the 
derivative operator is the best that can possibly be done for a 
given finite lattice spacing distance~$a$, the ``truncation'' of 
the operation to ``fit'' on a lattice with a finite number $N$ o%
f sites in [\ref{cit:Costella2002}]\ introduces small errors (of 
order $1/N$) that detract from the ``optimality'' of the operato%
r. It is desirable to remove these errors, to ensure that the st%
ochastic SLAC derivative operator for a \mbox{}\protect\/{%
\protect\em finite\protect\/} lattice is ``optimal''---namely, t%
he best that can possible be done.\par It is first worth seeing 
explicitly how the ``truncation'' operation detracts from the fi%
delity of the ideal SLAC derivative operator. Let us take a trul%
y extreme example---a lattice of only one site!---and let us ass%
ume that the underlying continuum formalism specifies that we ta%
ke the \mbox{}\protect\/{\protect\em second\protect\/}-derivativ%
e of some function $f(x)$. A simple derivation in [\ref{cit:Cost%
ella2002}]\ showed how one can obtain from first principles the 
expression first found by \protect\ref{au:Drell1976}~[\ref{cit:D%
rell1976}] for the ideal SLAC second-derivative operation in pos%
ition space, for the case of an infinite lattice: \setcounter{Ce%
qindent}{0}\protect\begin{eqnarray}\protect\left.\protect\begin{%
array}{rcl}\protect\displaystyle{\protect\it\Delta\!\:}^{(2)}_{%
\mbox{\scriptsize ideal}}f(x)\hspace{-1.3ex}&\protect
\displaystyle=&\hspace{-1.3ex}\protect\displaystyle\mbox{$%
\protect\displaystyle\protect\frac{2}{a^{2\!}}$}\setcounter{Cbsc%
urr}{25}\setlength{\Cscr}{\value{Cbscurr}\Ctenthex}\addtolength{%
\Cscr}{-1.0ex}\protect\raisebox{0ex}[\value{Cbscurr}\Ctenthex][%
\Cscr]{}\hspace{-0ex}{\protect\left\{\setlength{\Cscr}{\value{Cb%
scurr}\Ctenthex}\addtolength{\Cscr}{-1.0ex}\protect\raisebox{0ex%
}[\value{Cbscurr}\Ctenthex][\Cscr]{}\protect\right.}\hspace{-0.2%
5ex}\setlength{\Cscr}{\value{Cbscurr}\Ctenthex}\addtolength{\Cscr
}{-1.0ex}\protect\raisebox{0ex}[\value{Cbscurr}\Ctenthex][\Cscr]%
{}\setcounter{CbsD}{\value{CbsC}}\setcounter{CbsC}{\value{CbsB}}%
\setcounter{CbsB}{\value{CbsA}}\setcounter{CbsA}{\value{Cbscurr}%
}\ldots-\mbox{$\protect\displaystyle\protect\frac{1}{16}$}f(x+4a%
)+\mbox{$\protect\displaystyle\protect\frac{1}{9}$}f(x+3a)-\mbox
{$\protect\displaystyle\protect\frac{1}{4}$}f(x+2a)\setcounter{C%
eqindent}{185}\setlength{\Cscr}{\value{CEht}\Ctenthex}%
\addtolength{\Cscr}{-1.0ex}\protect\raisebox{0ex}[\value{CEht}%
\Ctenthex][\Cscr]{}\\*[0.55ex]\protect\displaystyle\hspace{-1.3e%
x}&\protect\displaystyle&\hspace{-1.3ex}\protect\displaystyle{%
\protect\mbox{}}\hspace{\value{Ceqindent}\Ctenthex}\setcounter{C%
Eht}{30}+f(x+a)-\mbox{$\protect\displaystyle\protect\frac{{\pi}{%
}^{\raisebox{-0.25ex}{$\scriptstyle2\!\!$}}}{6}$}f(x)+f(x-a)%
\setcounter{Ceqindent}{150}\setlength{\Cscr}{\value{CEht}%
\Ctenthex}\addtolength{\Cscr}{-1.0ex}\protect\raisebox{0ex}[%
\value{CEht}\Ctenthex][\Cscr]{}\\*[0.55ex]\protect\displaystyle
\hspace{-1.3ex}&\protect\displaystyle&\hspace{-1.3ex}\protect
\displaystyle{\protect\mbox{}}\hspace{\value{Ceqindent}\Ctenthex
}-\mbox{$\protect\displaystyle\protect\frac{1}{4}$}f(x-2a)+\mbox
{$\protect\displaystyle\protect\frac{1}{9}$}f(x-3a)-\mbox{$%
\protect\displaystyle\protect\frac{1}{16}$}f(x-4a)+\ldots
\setlength{\Cscr}{\value{CbsA}\Ctenthex}\addtolength{\Cscr}{-1.0%
ex}\protect\raisebox{0ex}[\value{CbsA}\Ctenthex][\Cscr]{}\hspace
{-0.25ex}{\protect\left.\setlength{\Cscr}{\value{CbsA}\Ctenthex}%
\addtolength{\Cscr}{-1.0ex}\protect\raisebox{0ex}[\value{CbsA}%
\Ctenthex][\Cscr]{}\protect\right\}}\hspace{-0ex}\setlength{\Cscr
}{\value{CbsA}\Ctenthex}\addtolength{\Cscr}{-1.0ex}\protect
\raisebox{0ex}[\value{CbsA}\Ctenthex][\Cscr]{}\setcounter{CbsA}{%
\value{CbsB}}\setcounter{CbsB}{\value{CbsC}}\setcounter{CbsC}{%
\value{CbsD}}\setcounter{CbsD}{1}.\setlength{\Cscr}{\value{CEht}%
\Ctenthex}\addtolength{\Cscr}{-1.0ex}\protect\raisebox{0ex}[%
\value{CEht}\Ctenthex][\Cscr]{}\protect\end{array}\protect\right
.\protect\label{eq:Truncation-De2Ideal}\protect\end{eqnarray}%
\setcounter{CEht}{10}What happens when we ``truncate'' this oper%
ation to apply to the extreme case of a lattice with just one si%
te? Clearly, we will be left with only the middle term, namely, 
\setcounter{Ceqindent}{0}\protect\begin{eqnarray}\protect\left.%
\protect\begin{array}{rcl}\protect\displaystyle\hspace{-1.3ex}&%
\protect\displaystyle{\protect\it\Delta\!\:}^{(2)}_{\mbox{%
\scriptsize truncated}}f(x)\setlength{\Cscr}{\value{Cbscurr}%
\Ctenthex}\addtolength{\Cscr}{-1.0ex}\protect\raisebox{0ex}[%
\value{Cbscurr}\Ctenthex][\Cscr]{}\hspace{-0.05ex}{\protect\left
|\setlength{\Cscr}{\value{Cbscurr}\Ctenthex}\addtolength{\Cscr}{%
-1.0ex}\protect\raisebox{0ex}[\value{Cbscurr}\Ctenthex][\Cscr]{}%
\protect\right.}\hspace{-0.25ex}\setlength{\Cscr}{\value{Cbscurr%
}\Ctenthex}\addtolength{\Cscr}{-1.0ex}\protect\raisebox{0ex}[%
\value{Cbscurr}\Ctenthex][\Cscr]{}_{x=0}\!\!=-\mbox{$\protect
\displaystyle\protect\frac{{\pi}{}^{\raisebox{-0.25ex}{$%
\scriptstyle2$}}}{3a^2}$}f(0).\setlength{\Cscr}{\value{CEht}%
\Ctenthex}\addtolength{\Cscr}{-1.0ex}\protect\raisebox{0ex}[%
\value{CEht}\Ctenthex][\Cscr]{}\protect\end{array}\protect\right
.\protect\label{eq:Truncation-TruncatedExtreme}\protect\end{eqna%
rray}\setcounter{CEht}{10}Assuming \mbox{$\protect\displaystyle
f(0)\neq0$}, we have somehow estimated the second-derivative of 
a function evaluated at just one point in space to be some nonze%
ro value! Of course, I said above that the errors in truncating 
the operator are of order $1/N$; and here \mbox{$\protect
\displaystyle N=1$}, so the errors are of order unity; in other 
words, the value itself has absolutely no reliability at all. Bu%
t is this the best we can do?\par I believe that it is not. When 
we restrict any physical formalism to a finite ``box''\mbox{$\!$%
}, we generally assume that the physical functions of relevance 
shall have periodic boundary conditions applied to them. (Indeed%
, it would be difficult to formulate a tractable analysis of mos%
t systems in momentum space if this stipulation were not to be m%
ade.) This is equivalent to thinking of a system of infinite spa%
tial extent, made up of ``boxes'' stacked side-by-side along the 
$x$-axis, with the extra proviso that the contents of each box a%
re to be considered to be identically equivalent to that of the 
``central box''\mbox{$\!$}. If this is the case, then in the ext%
reme case of a lattice with only one site, we are effectively st%
ipulating that the value of any function at every site of the im%
agined infinite lattice (one such site in each of these identica%
l ``boxes'') must be identical to the value of the function in t%
he ``central box'' (centred\ on \mbox{$\protect\displaystyle x=0%
$}). In other words, the function is effectively a constant on a%
ll of the (infinite number of) lattice sites: \setcounter{Ceqind%
ent}{0}\protect\begin{eqnarray}\hspace{-1.3ex}&\displaystyle f(x%
_{n\!}\equiv na)=f(0)=\mbox{constant}.\protect\nonumber\setlength
{\Cscr}{\value{CEht}\Ctenthex}\addtolength{\Cscr}{-1.0ex}\protect
\raisebox{0ex}[\value{CEht}\Ctenthex][\Cscr]{}\protect\end{eqnar%
ray}\setcounter{CEht}{10}In this context, our obtaining a nonzer%
o estimate of the second-derivative in (\protect\ref{eq:Truncati%
on-TruncatedExtreme}) looks very poor: surely, the second-deriva%
tive of a constant function is just zero!\par\refstepcounter{sec%
tion}\vspace{1.5\baselineskip}\par{\centering\bf\thesection. App%
lying periodic boundary conditions to the SLAC derivative operat%
ors\\*[0.5\baselineskip]}\protect\indent\label{sect:Periodic}The 
reason that the result (\protect\ref{eq:Truncation-TruncatedExtr%
eme}) obtained above is so poor is simply because we have (by as%
sumption) applied periodic boundary conditions to the function d%
efined on the single lattice site, but \mbox{}\protect\/{\protect
\em we have not applied periodic boundary conditions to the SLAC 
derivative operator\protect\/}; rather, we simply ``truncated'' 
it when we had gone once around the lattice (in this case, after 
evaluating it at just the single lattice point). Now, the SLAC s%
econd-derivative operator (\protect\ref{eq:Truncation-De2Ideal}) 
links the site in question to every other site on an infinite la%
ttice. Applying periodic boundary conditions to it implies that 
we must go around and around the lattice an infinite number of t%
imes, calculating all of these infinite number of terms that con%
tribute to the sum.\par Performing an infinite number of calcula%
tions does not sound very palatable, from a computational point 
of view. However, by the very assumption of periodic boundary co%
nditions, the function $f(x)$ itself is identically the same eve%
ry time we ``go past'' it, and so it effectively factorises out 
of the sum, namely, \setcounter{Ceqindent}{0}\protect\begin{eqna%
rray}\hspace{-1.3ex}&\displaystyle{\protect\it\Delta\!\:}^{(2)}_%
{\mbox{\scriptsize ideal}}f(x)\setlength{\Cscr}{\value{Cbscurr}%
\Ctenthex}\addtolength{\Cscr}{-1.0ex}\protect\raisebox{0ex}[%
\value{Cbscurr}\Ctenthex][\Cscr]{}\hspace{-0.05ex}{\protect\left
|\setlength{\Cscr}{\value{Cbscurr}\Ctenthex}\addtolength{\Cscr}{%
-1.0ex}\protect\raisebox{0ex}[\value{Cbscurr}\Ctenthex][\Cscr]{}%
\protect\right.}\hspace{-0.25ex}\setlength{\Cscr}{\value{Cbscurr%
}\Ctenthex}\addtolength{\Cscr}{-1.0ex}\protect\raisebox{0ex}[%
\value{Cbscurr}\Ctenthex][\Cscr]{}_{x=0}\!\!=c_0\:\!f(0),\protect
\nonumber\setlength{\Cscr}{\value{CEht}\Ctenthex}\addtolength{%
\Cscr}{-1.0ex}\protect\raisebox{0ex}[\value{CEht}\Ctenthex][\Cscr
]{}\protect\end{eqnarray}\setcounter{CEht}{10}where \setcounter{%
Ceqindent}{0}\protect\begin{eqnarray}\protect\left.\protect\begin
{array}{rcl}\protect\displaystyle\hspace{-1.3ex}&\protect
\displaystyle c_0\equiv\mbox{$\protect\displaystyle\protect\frac
{2}{a^2}$}\!\protect\left\{\ldots-\mbox{$\protect\displaystyle
\protect\frac{1}{16}$}+\mbox{$\protect\displaystyle\protect\frac
{1}{9}$}-\mbox{$\protect\displaystyle\protect\frac{1}{4}$}+1-%
\mbox{$\protect\displaystyle\protect\frac{{\pi}{}^{\raisebox{-0.%
25ex}{$\scriptstyle2\!\!$}}}{6}$}+1-\mbox{$\protect\displaystyle
\protect\frac{1}{4}$}+\mbox{$\protect\displaystyle\protect\frac{%
1}{9}$}-\mbox{$\protect\displaystyle\protect\frac{1}{16}$}+\ldots
\protect\right\}\!.\setlength{\Cscr}{\value{CEht}\Ctenthex}%
\addtolength{\Cscr}{-1.0ex}\protect\raisebox{0ex}[\value{CEht}%
\Ctenthex][\Cscr]{}\protect\end{array}\protect\right.\protect
\label{eq:Periodic-DefineC0}\protect\end{eqnarray}\setcounter{CE%
ht}{10}But this is simply an infinite sum of numerical factors, 
which needs to be only done once. Specifically, (\protect\ref{eq%
:Periodic-DefineC0}) is equivalent to \setcounter{Ceqindent}{0}%
\protect\begin{eqnarray}\hspace{-1.3ex}&\displaystyle c_0=\mbox{%
$\protect\displaystyle\protect\frac{2}{a^2}$}\!\protect\left\{2%
\sum_{n=1}^\infty\mbox{$\protect\displaystyle\protect\frac{(-1)^%
{n+1}}{n^2}$}-\mbox{$\protect\displaystyle\protect\frac{{\pi}{}^%
{\raisebox{-0.25ex}{$\scriptstyle2\!\!$}}}{6}$}\protect\right\}%
\!.\protect\nonumber\setlength{\Cscr}{\value{CEht}\Ctenthex}%
\addtolength{\Cscr}{-1.0ex}\protect\raisebox{0ex}[\value{CEht}%
\Ctenthex][\Cscr]{}\protect\end{eqnarray}\setcounter{CEht}{10}Fo%
rtunately, the mathematicians have worked out this infinite sum 
for us, namely, \setcounter{Ceqindent}{0}\protect\begin{eqnarray%
}\protect\left.\protect\begin{array}{rcl}\protect\displaystyle
\hspace{-1.3ex}&\protect\displaystyle\sum_{n=1}^\infty\mbox{$%
\protect\displaystyle\protect\frac{(-1)^{n+1}}{n^2}$}\equiv\eta(%
2)=\mbox{$\protect\displaystyle\protect\frac{{\pi}{}^{\raisebox{%
-0.25ex}{$\scriptstyle2\!\!$}}}{12}$}.\setlength{\Cscr}{\value{C%
Eht}\Ctenthex}\addtolength{\Cscr}{-1.0ex}\protect\raisebox{0ex}[%
\value{CEht}\Ctenthex][\Cscr]{}\protect\end{array}\protect\right
.\protect\label{eq:Periodic-Eta2}\protect\end{eqnarray}%
\setcounter{CEht}{10}Thus, we find \setcounter{Ceqindent}{0}%
\protect\begin{eqnarray}\hspace{-1.3ex}&\displaystyle c_0\equiv0%
,\protect\nonumber\setlength{\Cscr}{\value{CEht}\Ctenthex}%
\addtolength{\Cscr}{-1.0ex}\protect\raisebox{0ex}[\value{CEht}%
\Ctenthex][\Cscr]{}\protect\end{eqnarray}\setcounter{CEht}{10}an%
d hence the ideal SLAC second-derivative of any function defined 
on a lattice with just one site, with periodic boundary conditio%
ns applied to both the function \mbox{}\protect\/{\protect\em an%
d the operator\protect\/}, is identically zero, as argued above.%
\par\refstepcounter{section}\vspace{1.5\baselineskip}\par{%
\centering\bf\thesection. A more realistic example\\*[0.5%
\baselineskip]}\protect\indent\label{sect:Realistic}Let us now c%
onsider a more realistic example. For simplicity, let's consider 
a one-dimensional lattice with, say, \mbox{$\protect\displaystyle
N=100$} sites. Let us also go back to considering the \mbox{}%
\protect\/{\protect\em first\protect\/}-derivative of some funct%
ion $f(x)$ defined on the lattice. Now, since we are assuming pe%
riodic boundary conditions, it doesn't really matter how we labe%
l the 100 lattice sites relative to the site that we wish to com%
pute the derivative at; but, for definiteness, let us label them 
from \mbox{$\protect\displaystyle n=-49$} through \mbox{$\protect
\displaystyle n=+50$} inclusive, with \mbox{$\protect
\displaystyle n=0$} being the site that we wish to compute the f%
irst-derivative at. Now, the SLAC derivative operator (\protect
\ref{eq:Truncation-DeIdeal}) tells us to compute differences for 
increasing distances from \mbox{$\protect\displaystyle n=0$}, we%
ighting the difference at distance $n$ by \mbox{$\protect
\displaystyle(-1)^{n+1}\!/na$}. In [\ref{cit:Costella2002}], we 
simply assumed that we should stop when we get to the difference 
at \mbox{$\protect\displaystyle n=\pm49$}, since there is no 
\mbox{$\protect\displaystyle n=-50$}. Let us consider, as an arb%
itrary example, the lattice position at \mbox{$\protect
\displaystyle n=+43$}. The contribution to the sum (\protect\ref
{eq:Truncation-DeIdeal}) from this position, according to the tr%
uncated prescription of [\ref{cit:Costella2002}], is simply 
\setcounter{Ceqindent}{0}\protect\begin{eqnarray}\hspace{-1.3ex}%
&\displaystyle\mbox{$\protect\displaystyle\protect\frac{1}{43a}$%
}^{\:\!}f(x_{43}).\protect\nonumber\setlength{\Cscr}{\value{CEht%
}\Ctenthex}\addtolength{\Cscr}{-1.0ex}\protect\raisebox{0ex}[%
\value{CEht}\Ctenthex][\Cscr]{}\protect\end{eqnarray}\setcounter
{CEht}{10}\par We now wish to apply periodic boundary conditions 
to this derivative operation. Let us continue to concentrate on 
the position at \mbox{$\protect\displaystyle n=+43$}. When the d%
erivative operation ``runs'' through the lattice sites a second 
time, we will return to this same position, \mbox{$\protect
\displaystyle n=+43$}, after going a further distance of \mbox{$%
\protect\displaystyle N=100$} lattice sites. This implies a cont%
ribution to the sum in (\protect\ref{eq:Truncation-DeIdeal}) of 
the amount \setcounter{Ceqindent}{0}\protect\begin{eqnarray}%
\hspace{-1.3ex}&\displaystyle\mbox{$\protect\displaystyle\protect
\frac{1}{143a}$}^{\:\!}f(x_{143}\equiv x_{43}).\protect\nonumber
\setlength{\Cscr}{\value{CEht}\Ctenthex}\addtolength{\Cscr}{-1.0%
ex}\protect\raisebox{0ex}[\value{CEht}\Ctenthex][\Cscr]{}\protect
\end{eqnarray}\setcounter{CEht}{10}After another ``run'' through 
the lattice sites, we will add a further contribution of 
\setcounter{Ceqindent}{0}\protect\begin{eqnarray}\hspace{-1.3ex}%
&\displaystyle\mbox{$\protect\displaystyle\protect\frac{1}{243a}%
$}^{\:\!}f(x_{43}).\protect\nonumber\setlength{\Cscr}{\value{CEh%
t}\Ctenthex}\addtolength{\Cscr}{-1.0ex}\protect\raisebox{0ex}[%
\value{CEht}\Ctenthex][\Cscr]{}\protect\end{eqnarray}\setcounter
{CEht}{10}Thus, continuing on this process \mbox{}\protect\/{%
\protect\em ad infinitum\protect\/}, the overall contribution to 
the sum in (\protect\ref{eq:Truncation-DeIdeal}) will simply be 
\setcounter{Ceqindent}{0}\protect\begin{eqnarray}\hspace{-1.3ex}%
&\displaystyle\mbox{$\protect\displaystyle\protect\frac{1}{a}$}%
\!\protect\left\{\mbox{$\protect\displaystyle\protect\frac{1}{43%
}$}+\mbox{$\protect\displaystyle\protect\frac{1}{143}$}+\mbox{$%
\protect\displaystyle\protect\frac{1}{243}$}+\mbox{$\protect
\displaystyle\protect\frac{1}{343}$}+\ldots\protect\right\}\!f(x%
_{43}).\protect\nonumber\setlength{\Cscr}{\value{CEht}\Ctenthex}%
\addtolength{\Cscr}{-1.0ex}\protect\raisebox{0ex}[\value{CEht}%
\Ctenthex][\Cscr]{}\protect\end{eqnarray}\setcounter{CEht}{10}It 
might seem that the coefficient of this term is simply the coeff%
icient $c_{43}$ that we seek to evaluate. However, we have a pro%
blem: this infinite sum is \mbox{}\protect\/{\protect\em diverge%
nt\protect\/}, as can be seen from the fact that \mbox{$\protect
\displaystyle1/43>1/100$}, \mbox{$\protect\displaystyle1/143>1/2%
00$}, \mbox{$\protect\displaystyle1/243>1/300$}, and so on, so t%
hat \setcounter{Ceqindent}{0}\protect\begin{eqnarray}\mbox{$%
\protect\displaystyle\protect\frac{1}{43}$}+\mbox{$\protect
\displaystyle\protect\frac{1}{143}$}+\mbox{$\protect\displaystyle
\protect\frac{1}{243}$}+\mbox{$\protect\displaystyle\protect\frac
{1}{343}$}+\ldots\hspace{-1.3ex}&\displaystyle>&\hspace{-1.3ex}%
\mbox{$\protect\displaystyle\protect\frac{1}{100}$}+\mbox{$%
\protect\displaystyle\protect\frac{1}{200}$}+\mbox{$\protect
\displaystyle\protect\frac{1}{300}$}+\mbox{$\protect\displaystyle
\protect\frac{1}{400}$}+\ldots\protect\nonumber\setlength{\Cscr}%
{\value{CEht}\Ctenthex}\addtolength{\Cscr}{-1.0ex}\protect
\raisebox{0ex}[\value{CEht}\Ctenthex][\Cscr]{}\\*[0ex]\protect
\displaystyle\hspace{-1.3ex}&\displaystyle=&\hspace{-1.3ex}\mbox
{$\protect\displaystyle\protect\frac{1}{100}$}\!\protect\left\{1%
+\mbox{$\protect\displaystyle\protect\frac{1}{2}$}+\mbox{$%
\protect\displaystyle\protect\frac{1}{3}$}+\mbox{$\protect
\displaystyle\protect\frac{1}{4}$}+\ldots\protect\right\}\!,%
\protect\nonumber\setlength{\Cscr}{\value{CEht}\Ctenthex}%
\addtolength{\Cscr}{-1.0ex}\protect\raisebox{0ex}[\value{CEht}%
\Ctenthex][\Cscr]{}\protect\end{eqnarray}\setcounter{CEht}{10}an%
d this last sum is logarithmically divergent.\par What has gone 
wrong? Obviously, the correct coefficient cannot be infinite. Th%
e problem is that we have only considered one half of the deriva%
tive operation expressed in (\protect\ref{eq:Truncation-DeIdeal}%
), namely, that for positive~$n$. We have forgotten that, in tra%
velling backwards along the lattice for \mbox{}\protect\/{%
\protect\em negative\protect\/}~$n$, and applying periodic bound%
ary conditions at the boundary which we are taking to occur afte%
r \mbox{$\protect\displaystyle n=-49$}, we will also ``run past%
'' the position at \mbox{$\protect\displaystyle n=+43$}. Specifi%
cally, the position \mbox{$\protect\displaystyle n=-57$} is, by 
the periodic boundary conditions, identically equivalent to \mbox
{$\protect\displaystyle n=+43$}, and this term in (\protect\ref{%
eq:Truncation-DeIdeal}) will contribute an amount \setcounter{Ce%
qindent}{0}\protect\begin{eqnarray}\hspace{-1.3ex}&\displaystyle
-\mbox{$\protect\displaystyle\protect\frac{1}{57a}$}^{\:\!}f(x_{%
-57}\equiv x_{43}).\protect\nonumber\setlength{\Cscr}{\value{CEh%
t}\Ctenthex}\addtolength{\Cscr}{-1.0ex}\protect\raisebox{0ex}[%
\value{CEht}\Ctenthex][\Cscr]{}\protect\end{eqnarray}\setcounter
{CEht}{10}Going back another \mbox{$\protect\displaystyle N=100$%
} lattice sites, we will also find a contribution of \setcounter
{Ceqindent}{0}\protect\begin{eqnarray}\hspace{-1.3ex}&%
\displaystyle-\mbox{$\protect\displaystyle\protect\frac{1}{157a}%
$}^{\:\!}f(x_{-157}\equiv x_{43}),\protect\nonumber\setlength{%
\Cscr}{\value{CEht}\Ctenthex}\addtolength{\Cscr}{-1.0ex}\protect
\raisebox{0ex}[\value{CEht}\Ctenthex][\Cscr]{}\protect\end{eqnar%
ray}\setcounter{CEht}{10}and so on. Thus, we find that the true 
expression for $c_{43}$ is simply \setcounter{Ceqindent}{0}%
\protect\begin{eqnarray}\protect\left.\protect\begin{array}{rcl}%
\protect\displaystyle\hspace{-1.3ex}&\protect\displaystyle c_{43%
}=\mbox{$\protect\displaystyle\protect\frac{1}{a}$}\!\protect
\left\{\mbox{$\protect\displaystyle\protect\frac{1}{43}$}-\mbox{%
$\protect\displaystyle\protect\frac{1}{57}$}+\mbox{$\protect
\displaystyle\protect\frac{1}{143}$}-\mbox{$\protect\displaystyle
\protect\frac{1}{157}$}+\mbox{$\protect\displaystyle\protect\frac
{1}{243}$}-\mbox{$\protect\displaystyle\protect\frac{1}{257}$}+%
\ldots\protect\right\}\!.\setlength{\Cscr}{\value{CEht}\Ctenthex
}\addtolength{\Cscr}{-1.0ex}\protect\raisebox{0ex}[\value{CEht}%
\Ctenthex][\Cscr]{}\protect\end{array}\protect\right.\protect
\label{eq:Realistic-C43N100}\protect\end{eqnarray}\setcounter{CE%
ht}{10}It is straightforward to see that this alternating series 
is now convergent; for example, by noting that \mbox{$\protect
\displaystyle1/43<1/25$}, \mbox{$\protect\displaystyle1/57<1/50$%
}, \mbox{$\protect\displaystyle1/143<1/75$}, and so on, so that 
\setcounter{Ceqindent}{0}\protect\begin{eqnarray}\hspace{-1.3ex}%
&\displaystyle c_{43}<\mbox{$\protect\displaystyle\protect\frac{%
1}{a}$}\!\protect\left\{\mbox{$\protect\displaystyle\protect\frac
{1}{25}$}-\mbox{$\protect\displaystyle\protect\frac{1}{50}$}+%
\mbox{$\protect\displaystyle\protect\frac{1}{75}$}-\mbox{$%
\protect\displaystyle\protect\frac{1}{100}$}+\ldots\protect\right
\}=\mbox{$\protect\displaystyle\protect\frac{1}{25a}$}\!\protect
\left\{1-\mbox{$\protect\displaystyle\protect\frac{1}{2}$}+\mbox
{$\protect\displaystyle\protect\frac{1}{3}$}-\mbox{$\protect
\displaystyle\protect\frac{1}{4}$}+\ldots\protect\right\}=\mbox{%
$\protect\displaystyle\protect\frac{\ln2}{25a}$}<\infty.\protect
\nonumber\setlength{\Cscr}{\value{CEht}\Ctenthex}\addtolength{%
\Cscr}{-1.0ex}\protect\raisebox{0ex}[\value{CEht}\Ctenthex][\Cscr
]{}\protect\end{eqnarray}\setcounter{CEht}{10}\par Now, you may 
have noted that, in ``wrapping around'' the \mbox{$\protect
\displaystyle N=100$} lattice sites in the above, the fact that 
$N$ was \mbox{}\protect\/{\protect\em even\protect\/} meant that 
the sign of the contribution to (\protect\ref{eq:Truncation-DeId%
eal}) in any ``run'' through the lattice (in one direction) was 
the same as any other ``run'' through the lattice (in that same 
direction). How would the situation change if $N$ were \mbox{}%
\protect\/{\protect\em odd\protect\/}? To examine this situation%
, let us consider the example of \mbox{$\protect\displaystyle N=%
101$}, and let us continue to concentrate our attention on the p%
osition \mbox{$\protect\displaystyle n=+43$}. Going through the 
positive-$n$ terms in (\protect\ref{eq:Truncation-DeIdeal}), we 
will pick up contributions of \setcounter{CElett}{0}\protect
\refstepcounter{equation}\protect\label{eq:Realistic-C43N101}%
\renewcommand\theequation{\arabic{equation}\alph{CElett}}\protect
\stepcounter{CElett}\addtocounter{equation}{-1}\setcounter{Ceqin%
dent}{0}\protect\begin{eqnarray}\protect\left.\protect\begin{arr%
ay}{rcl}\protect\displaystyle\hspace{-1.3ex}&\protect
\displaystyle\mbox{$\protect\displaystyle\protect\frac{1}{a}$}\!%
\protect\left\{\mbox{$\protect\displaystyle\protect\frac{1}{43}$%
}-\mbox{$\protect\displaystyle\protect\frac{1}{144}$}+\mbox{$%
\protect\displaystyle\protect\frac{1}{245}$}-\mbox{$\protect
\displaystyle\protect\frac{1}{346}$}+\ldots\protect\right\}\mbox
{$\!$},\setlength{\Cscr}{\value{CEht}\Ctenthex}\addtolength{\Cscr
}{-1.0ex}\protect\raisebox{0ex}[\value{CEht}\Ctenthex][\Cscr]{}%
\protect\end{array}\protect\right.\protect\label{eq:Realistic-C4%
3N101Pos}\protect\end{eqnarray}\setcounter{CEht}{10}where the mi%
nus signs arise because the terms in (\protect\ref{eq:Truncation%
-DeIdeal}) for \mbox{}\protect\/{\protect\em even\protect\/} pos%
itive~$n$ come in with the minus signs. This sum is now, by itse%
lf, convergent. However, to obtain $c_{43}$ we still need to add 
in the contributions from the negative-$n$ terms in (\protect\ref
{eq:Truncation-DeIdeal}). The periodic boundary conditions now t%
ell us that the position \mbox{$\protect\displaystyle n=-58$} is 
equivalent to \mbox{$\protect\displaystyle n=43$} (because \mbox
{$\protect\displaystyle43+58=101$}), and the coefficient of \mbox
{$\protect\displaystyle f(x-58a)$} in (\protect\ref{eq:Truncatio%
n-DeIdeal}) is positive, so we will obtain the contribution 
\protect\stepcounter{CElett}\addtocounter{equation}{-1}%
\setcounter{Ceqindent}{0}\protect\begin{eqnarray}\protect\left.%
\protect\begin{array}{rcl}\protect\displaystyle\hspace{-1.3ex}&%
\protect\displaystyle\mbox{$\protect\displaystyle\protect\frac{1%
}{a}$}\!\protect\left\{\mbox{$\protect\displaystyle\protect\frac
{1}{58}$}-\mbox{$\protect\displaystyle\protect\frac{1}{159}$}+%
\mbox{$\protect\displaystyle\protect\frac{1}{260}$}-\mbox{$%
\protect\displaystyle\protect\frac{1}{261}$}+\ldots\protect\right
\}\mbox{$\!$},\setlength{\Cscr}{\value{CEht}\Ctenthex}%
\addtolength{\Cscr}{-1.0ex}\protect\raisebox{0ex}[\value{CEht}%
\Ctenthex][\Cscr]{}\protect\end{array}\protect\right.\protect
\label{eq:Realistic-C43N101Neg}\protect\end{eqnarray}\setcounter
{CEht}{10}\renewcommand\theequation{\arabic{equation}}which is a%
lso, by itself, convergent.\par\refstepcounter{section}\vspace{1%
.5\baselineskip}\par{\centering\bf\thesection. An example of the 
second derivative\\*[0.5\baselineskip]}\protect\indent\label{sec%
t:SecondEx}The same process as outlined in the example above can 
be applied to the second-derivative operator (\protect\ref{eq:Tr%
uncation-De2Ideal}) (and, indeed, any higher-order derivative th%
at may be desired). Let us go back to the example of \mbox{$%
\protect\displaystyle N=100$}, and consider again the position 
\mbox{$\protect\displaystyle n=+43$}. Under periodic boundary co%
nditions, the terms in (\protect\ref{eq:Truncation-De2Ideal}) fo%
r positive $n$ that contribution to this position are just 
\setcounter{CElett}{0}\protect\refstepcounter{equation}\protect
\label{eq:SecondEx-Even}\renewcommand\theequation{\arabic{equati%
on}\alph{CElett}}\protect\stepcounter{CElett}\addtocounter{equat%
ion}{-1}\setcounter{Ceqindent}{0}\protect\begin{eqnarray}\protect
\left.\protect\begin{array}{rcl}\protect\displaystyle\hspace{-1.%
3ex}&\protect\displaystyle\mbox{$\protect\displaystyle\protect
\frac{2}{a^{2\!}}$}\!\protect\left\{\mbox{$\protect\displaystyle
\protect\frac{1}{43^2}$}+\mbox{$\protect\displaystyle\protect
\frac{1}{143^2}$}+\mbox{$\protect\displaystyle\protect\frac{1}{2%
43^2}$}+\mbox{$\protect\displaystyle\protect\frac{1}{343^2}$}+%
\ldots\protect\right\}\!f(x_{43}).\setlength{\Cscr}{\value{CEht}%
\Ctenthex}\addtolength{\Cscr}{-1.0ex}\protect\raisebox{0ex}[%
\value{CEht}\Ctenthex][\Cscr]{}\protect\end{array}\protect\right
.\protect\label{eq:SecondEx-EvenPos}\protect\end{eqnarray}%
\setcounter{CEht}{10}Because these denominators are squared, rat%
her than linear, this sum converges on its own. The negative-$n$ 
terms in (\protect\ref{eq:Truncation-De2Ideal}) that contribute 
to this position are just \protect\stepcounter{CElett}%
\addtocounter{equation}{-1}\setcounter{Ceqindent}{0}\protect
\begin{eqnarray}\protect\left.\protect\begin{array}{rcl}\protect
\displaystyle\hspace{-1.3ex}&\protect\displaystyle\mbox{$\protect
\displaystyle\protect\frac{2}{a^{2\!}}$}\!\protect\left\{\mbox{$%
\protect\displaystyle\protect\frac{1}{57^2}$}+\mbox{$\protect
\displaystyle\protect\frac{1}{157^2}$}+\mbox{$\protect
\displaystyle\protect\frac{1}{257^2}$}+\mbox{$\protect
\displaystyle\protect\frac{1}{357^2}$}+\ldots\protect\right\}\!f%
(x_{43}),\setlength{\Cscr}{\value{CEht}\Ctenthex}\addtolength{%
\Cscr}{-1.0ex}\protect\raisebox{0ex}[\value{CEht}\Ctenthex][\Cscr
]{}\protect\end{array}\protect\right.\protect\label{eq:SecondEx-%
EvenNeg}\protect\end{eqnarray}\setcounter{CEht}{10}\renewcommand
\theequation{\arabic{equation}}which also converges. For even $N%
$, these two sums contribute with the same sign.\par It is clear 
that the same of calculation would also apply for the case of 
\mbox{$\protect\displaystyle n=+50$}, which is ``half a lattice 
away'' from \mbox{$\protect\displaystyle n=0$}, even though ther%
e is no corresponding \mbox{$\protect\displaystyle n=-50$} (whic%
h is, by the periodic boundary conditions, equivalent to \mbox{$%
\protect\displaystyle n=+50$}); for this position, the two sums 
of the form (\protect\ref{eq:SecondEx-EvenPos}) and (\protect\ref
{eq:SecondEx-EvenNeg}) will be identical.\par Now, because the c%
entral (\mbox{$\protect\displaystyle n=0$}) term in the operator 
(\protect\ref{eq:Truncation-De2Ideal}) is nonzero, and because e%
ach term for $-n$ comes in with the \mbox{}\protect\/{\protect\em
same\protect\/} sign as that for $+n$, we also need to separatel%
y compute the coefficient $c_0$ under the assumption of periodic 
boundary conditions. For the case of \mbox{$\protect\displaystyle
N=100$}, we simply obtain \setcounter{Ceqindent}{0}\protect\begin
{eqnarray}\protect\left.\protect\begin{array}{rcl}\protect
\displaystyle\hspace{-1.3ex}&\protect\displaystyle c_0=\mbox{$%
\protect\displaystyle\protect\frac{2}{a^{2\!}}$}\!\protect\left
\{-\mbox{$\protect\displaystyle\protect\frac{{\pi}{}^{\raisebox{%
-0.25ex}{$\scriptstyle2\!$}}}{6}$}-\mbox{$\protect\displaystyle
\protect\frac{2}{100^2}$}-\mbox{$\protect\displaystyle\protect
\frac{2}{200^2}$}-\mbox{$\protect\displaystyle\protect\frac{2}{3%
00^2}$}-\ldots\protect\right\}\!,\setlength{\Cscr}{\value{CEht}%
\Ctenthex}\addtolength{\Cscr}{-1.0ex}\protect\raisebox{0ex}[%
\value{CEht}\Ctenthex][\Cscr]{}\protect\end{array}\protect\right
.\protect\label{eq:SecondEx-EvenZero}\protect\end{eqnarray}%
\setcounter{CEht}{10}where the factor of 2 in each term arises f%
rom the fact that we are adding in the contributions from both p%
ositive-$n$ and negative-$n$ (which are both one lattice ``revol%
ution'' away).\par Let us now go back to the odd-$N$ example of 
\mbox{$\protect\displaystyle N=101$}. For the position \mbox{$%
\protect\displaystyle n=+43$}, the positive-$n$ terms that contr%
ibute are just \setcounter{CElett}{0}\protect\refstepcounter{equ%
ation}\protect\label{eq:SecondEx-Odd}\renewcommand\theequation{%
\arabic{equation}\alph{CElett}}\protect\stepcounter{CElett}%
\addtocounter{equation}{-1}\setcounter{Ceqindent}{0}\protect
\begin{eqnarray}\protect\left.\protect\begin{array}{rcl}\protect
\displaystyle\hspace{-1.3ex}&\protect\displaystyle\mbox{$\protect
\displaystyle\protect\frac{2}{a^{2\!}}$}\!\protect\left\{\mbox{$%
\protect\displaystyle\protect\frac{1}{43^2}$}-\mbox{$\protect
\displaystyle\protect\frac{1}{144^2}$}+\mbox{$\protect
\displaystyle\protect\frac{1}{245^2}$}-\mbox{$\protect
\displaystyle\protect\frac{1}{346^2}$}+\ldots\protect\right\}\!f%
(x_{43}),\setlength{\Cscr}{\value{CEht}\Ctenthex}\addtolength{%
\Cscr}{-1.0ex}\protect\raisebox{0ex}[\value{CEht}\Ctenthex][\Cscr
]{}\protect\end{array}\protect\right.\protect\label{eq:SecondEx-%
OddPos}\protect\end{eqnarray}\setcounter{CEht}{10}and the negati%
ve-$n$ terms that contribute are just \protect\stepcounter{CElet%
t}\addtocounter{equation}{-1}\setcounter{Ceqindent}{0}\protect
\begin{eqnarray}\protect\left.\protect\begin{array}{rcl}\protect
\displaystyle\hspace{-1.3ex}&\protect\displaystyle\mbox{$\protect
\displaystyle\protect\frac{2}{a^{2\!}}$}\!\protect\left\{-\mbox{%
$\protect\displaystyle\protect\frac{1}{58^2}$}+\mbox{$\protect
\displaystyle\protect\frac{1}{159^2}$}-\mbox{$\protect
\displaystyle\protect\frac{1}{260^2}$}+\mbox{$\protect
\displaystyle\protect\frac{1}{361^{\:\!\!2}}$}+\ldots\protect
\right\}\!f(x_{43}).\setlength{\Cscr}{\value{CEht}\Ctenthex}%
\addtolength{\Cscr}{-1.0ex}\protect\raisebox{0ex}[\value{CEht}%
\Ctenthex][\Cscr]{}\protect\end{array}\protect\right.\protect
\label{eq:SecondEx-OddNeg}\protect\end{eqnarray}\setcounter{CEht%
}{10}\renewcommand\theequation{\arabic{equation}}The extra minus 
signs that arise from using an odd number of lattice sites make 
the overall sum converge more quickly than for an even value of~%
$N$. For the ``central'' coefficient $c_0$, we now obtain 
\setcounter{Ceqindent}{0}\protect\begin{eqnarray}\protect\left.%
\protect\begin{array}{rcl}\protect\displaystyle\hspace{-1.3ex}&%
\protect\displaystyle c_0=\mbox{$\protect\displaystyle\protect
\frac{2}{a^{2\!}}$}\!\protect\left\{-\mbox{$\protect\displaystyle
\protect\frac{{\pi}{}^{\raisebox{-0.25ex}{$\scriptstyle2\!\!$}}}%
{6}$}+\mbox{$\protect\displaystyle\protect\frac{2}{101^{\:\!\!2}%
}$}-\mbox{$\protect\displaystyle\protect\frac{2}{202^2}$}+\mbox{%
$\protect\displaystyle\protect\frac{2}{303^2}$}-\ldots\protect
\right\}\!,\setlength{\Cscr}{\value{CEht}\Ctenthex}\addtolength{%
\Cscr}{-1.0ex}\protect\raisebox{0ex}[\value{CEht}\Ctenthex][\Cscr
]{}\protect\end{array}\protect\right.\protect\label{eq:SecondEx-%
OddZero}\protect\end{eqnarray}\setcounter{CEht}{10}where the min%
us signs are again due to the fact that we are alternately picki%
ng up the odd and even terms in (\protect\ref{eq:Truncation-De2I%
deal}).\par\refstepcounter{section}\vspace{1.5\baselineskip}\par
{\centering\bf\thesection. Exact, general expressions for the fi%
nite-lattice SLAC derivative operators\\*[0.5\baselineskip]}%
\protect\indent\label{sect:Exact}Let us now make use of the exam%
ples provided in \mbox{Secs.~$\:\!\!$}\protect\ref{sect:Realisti%
c} and~\protect\ref{sect:SecondEx} to guide us in constructing, 
from first principles, the general expressions for the SLAC deri%
vative operators on a finite lattice on which we assume periodic 
boundary conditions.\par Let us start with the first-derivative 
operator. We can write the expression (\protect\ref{eq:Realistic%
-C43N100}) for $c_{43}$ with \mbox{$\protect\displaystyle N=100$%
} in the form \setcounter{Ceqindent}{0}\protect\begin{eqnarray}%
\hspace{-1.3ex}&\displaystyle{}^{(1)\!}c_{43}^{100}=\mbox{$%
\protect\displaystyle\protect\frac{1}{a}$}\sum_{k=0}^\infty\!%
\protect\left\{\mbox{$\protect\displaystyle\protect\frac{1}{100k%
+43}$}-\mbox{$\protect\displaystyle\protect\frac{1}{100k+57}$}%
\protect\right\}\!,\protect\nonumber\setlength{\Cscr}{\value{CEh%
t}\Ctenthex}\addtolength{\Cscr}{-1.0ex}\protect\raisebox{0ex}[%
\value{CEht}\Ctenthex][\Cscr]{}\protect\end{eqnarray}\setcounter
{CEht}{10}where I am using the notation \setcounter{Ceqindent}{0%
}\protect\begin{eqnarray}\hspace{-1.3ex}&\displaystyle{}^{(m)\!}%
c_n^N\protect\nonumber\setlength{\Cscr}{\value{CEht}\Ctenthex}%
\addtolength{\Cscr}{-1.0ex}\protect\raisebox{0ex}[\value{CEht}%
\Ctenthex][\Cscr]{}\protect\end{eqnarray}\setcounter{CEht}{10}fo%
r the coefficient of the ideal SLAC $m$-th derivative operator a%
t position $n$ for a finite lattice of $N$ lattice sites. If we 
had chosen a general lattice position $n$ (positive) rather than 
the particular choice of \mbox{$\protect\displaystyle n=43$}, we 
would have obtained \setcounter{Ceqindent}{0}\protect\begin{eqna%
rray}\hspace{-1.3ex}&\displaystyle{}^{(1)\!}c_{\mbox{\scriptsize
$n$ positive}}^{100}=\mbox{$\protect\displaystyle\protect\frac{(%
-1)^{n+1}}{a}$}\sum_{k=0}^\infty\!\protect\left\{\mbox{$\protect
\displaystyle\protect\frac{1}{100k+n}$}-\mbox{$\protect
\displaystyle\protect\frac{1}{100k+100-n}$}\protect\right\}\!,%
\protect\nonumber\setlength{\Cscr}{\value{CEht}\Ctenthex}%
\addtolength{\Cscr}{-1.0ex}\protect\raisebox{0ex}[\value{CEht}%
\Ctenthex][\Cscr]{}\protect\end{eqnarray}\setcounter{CEht}{10}wh%
ere the alternating sign out the front is due to the alternating 
sign in (\protect\ref{eq:Truncation-DeIdeal}). Clearly, if we we%
re to have a general (even) number $N$ of lattice sites, this re%
sult would become \setcounter{Ceqindent}{0}\protect\begin{eqnarr%
ay}\hspace{-1.3ex}&\displaystyle{}^{(1)\!}c_{\mbox{\scriptsize$n%
$ positive}}^{\mbox{\scriptsize$N\!$ even}}=\mbox{$\protect
\displaystyle\protect\frac{(-1)^{n+1}}{a}$}\sum_{k=0}^\infty\!%
\protect\left\{\mbox{$\protect\displaystyle\protect\frac{1}{Nk+n%
}$}-\mbox{$\protect\displaystyle\protect\frac{1}{Nk+N-n}$}%
\protect\right\}\!.\protect\nonumber\setlength{\Cscr}{\value{CEh%
t}\Ctenthex}\addtolength{\Cscr}{-1.0ex}\protect\raisebox{0ex}[%
\value{CEht}\Ctenthex][\Cscr]{}\protect\end{eqnarray}\setcounter
{CEht}{10}If we factorise out a factor of $N$ from each of these 
denominators, we obtain \setcounter{Ceqindent}{0}\protect\begin{%
eqnarray}\hspace{-1.3ex}&\displaystyle{}^{(1)\!}c_{\mbox{%
\scriptsize$n$ positive}}^{\mbox{\scriptsize$N\!$ even}}=\mbox{$%
\protect\displaystyle\protect\frac{(-1)^{n+1}}{Na}$}\sum_{k=0}^%
\infty\!\protect\left\{\mbox{$\protect\displaystyle\protect\frac
{1}{k+n/N}$}-\mbox{$\protect\displaystyle\protect\frac{1}{k+1-n/%
N}$}\protect\right\}\!.\protect\nonumber\setlength{\Cscr}{\value
{CEht}\Ctenthex}\addtolength{\Cscr}{-1.0ex}\protect\raisebox{0ex%
}[\value{CEht}\Ctenthex][\Cscr]{}\protect\end{eqnarray}%
\setcounter{CEht}{10}Fortunately, the mathematicians have again 
evaluated this infinite sum for us, and it yields a remarkably s%
imple result: \setcounter{Ceqindent}{0}\protect\begin{eqnarray}%
\protect\left.\protect\begin{array}{rcl}\protect\displaystyle
\hspace{-1.3ex}&\protect\displaystyle\sum_{k=0}^\infty\!\protect
\left\{\mbox{$\protect\displaystyle\protect\frac{1}{k+n/N}$}-%
\mbox{$\protect\displaystyle\protect\frac{1}{k+1-n/N}$}\protect
\right\}=\pi\cot\!\protect\left(\mbox{$\protect\displaystyle
\protect\frac{\pi n}{N}$}\protect\right)\!.\setlength{\Cscr}{%
\value{CEht}\Ctenthex}\addtolength{\Cscr}{-1.0ex}\protect
\raisebox{0ex}[\value{CEht}\Ctenthex][\Cscr]{}\protect\end{array%
}\protect\right.\protect\label{eq:Exact-InfiniteSumLinears1}%
\protect\end{eqnarray}\setcounter{CEht}{10}Thus, we find that 
\setcounter{Ceqindent}{0}\protect\begin{eqnarray}\hspace{-1.3ex}%
&\displaystyle{}^{(1)\!}c_{\mbox{\scriptsize$n$ positive}}^{\mbox
{\scriptsize$N\!$ even}}=(-1)^{n+1}\mbox{$\protect\displaystyle
\protect\frac{\pi}{Na}$}\cot\!\protect\left(\mbox{$\protect
\displaystyle\protect\frac{\pi n}{N}$}\protect\right)\!.\protect
\nonumber\setlength{\Cscr}{\value{CEht}\Ctenthex}\addtolength{%
\Cscr}{-1.0ex}\protect\raisebox{0ex}[\value{CEht}\Ctenthex][\Cscr
]{}\protect\end{eqnarray}\setcounter{CEht}{10}Let us now conside%
r a negative value of $n$ (but keeping $N$ even). {}From\ (%
\protect\ref{eq:Truncation-DeIdeal}), all of the terms come in w%
ith the opposite sign, but since \mbox{$\protect\displaystyle(-1%
)^{-n+1}\equiv(-1)^{n+1}$} and \mbox{$\protect\displaystyle\cot(%
-x)\equiv-\cot x$}, we can see that the same expression can be u%
sed for negative~$n$. Hence, in general (for even~$N$), we have 
\setcounter{Ceqindent}{0}\protect\begin{eqnarray}\protect\left.%
\protect\begin{array}{rcl}\protect\displaystyle\hspace{-1.3ex}&%
\protect\displaystyle{}^{(1)\!}c_{n\neq0}^{\mbox{\scriptsize$N\!%
$ even}}=(-1)^{n+1}\mbox{$\protect\displaystyle\protect\frac{\pi
}{Na}$}\cot\!\protect\left(\mbox{$\protect\displaystyle\protect
\frac{\pi n}{N}$}\protect\right)\!.\setlength{\Cscr}{\value{CEht%
}\Ctenthex}\addtolength{\Cscr}{-1.0ex}\protect\raisebox{0ex}[%
\value{CEht}\Ctenthex][\Cscr]{}\protect\end{array}\protect\right
.\protect\label{eq:Exact-C1EvenN}\protect\end{eqnarray}%
\setcounter{CEht}{10}It is clear that, for small $n$ (namely, fa%
r from the application of the periodic boundary conditions), thi%
s formula approaches the infinite lattice result (\protect\ref{e%
q:Truncation-DeIdeal}), because \mbox{$\protect\displaystyle\cot
x\equiv1/\tan x\sim1/x$} for small~$x$, and hence \setcounter{Ce%
qindent}{0}\protect\begin{eqnarray}\hspace{-1.3ex}&\displaystyle
{}^{(1)\!}c_{n\neq0}^{\mbox{\scriptsize$N\!$ even}}\rightarrow
\mbox{$\protect\displaystyle\protect\frac{(-1)^{n+1}}{na}$}%
\hspace{3ex}\mbox{as}\hspace{3ex}\mbox{$\protect\displaystyle
\protect\frac{n}{N}$}\rightarrow0.\protect\nonumber\setlength{%
\Cscr}{\value{CEht}\Ctenthex}\addtolength{\Cscr}{-1.0ex}\protect
\raisebox{0ex}[\value{CEht}\Ctenthex][\Cscr]{}\protect\end{eqnar%
ray}\setcounter{CEht}{10}On the other hand, near the edge of the 
application of periodic boundary conditions, namely, \mbox{$%
\protect\displaystyle n\rightarrow N/2$}, the absolute values of 
the coefficients (\protect\ref{eq:Exact-C1EvenN}) approach zero 
linearly with~\mbox{$\protect\displaystyle(N/2-n)$}.\par Let us 
now consider the case of an odd number of lattice sites~$N$. We 
again start at a position of positive~$n$. The example (\protect
\ref{eq:Realistic-C43N101Pos}) shows us that the positive-$n$ te%
rms in (\protect\ref{eq:Truncation-DeIdeal}) themselves contribu%
te an amount that can be written \setcounter{Ceqindent}{0}%
\protect\begin{eqnarray}\hspace{-1.3ex}&\displaystyle\mbox{$%
\protect\displaystyle\protect\frac{1}{a}$}\lim_{M\rightarrow
\infty}\!\protect\left\{2\sum_{k=0}^M\mbox{$\protect\displaystyle
\protect\frac{1}{202k+43}$}-\sum_{k=0}^{2M}\mbox{$\protect
\displaystyle\protect\frac{1}{101k+43}$}\protect\right\}\!,%
\protect\nonumber\setlength{\Cscr}{\value{CEht}\Ctenthex}%
\addtolength{\Cscr}{-1.0ex}\protect\raisebox{0ex}[\value{CEht}%
\Ctenthex][\Cscr]{}\protect\end{eqnarray}\setcounter{CEht}{10}wh%
ere we have written the two sums in this way (the first sum cont%
ributing to every \mbox{}\protect\/{\protect\em other\protect\/} 
term in (\protect\ref{eq:Realistic-C43N101Pos}), with the second 
sum contributing to \mbox{}\protect\/{\protect\em every\protect
\/} term) to maintain the closest possible relationship with the 
even-$N$ analysis above. (The limiting process is necessary beca%
use each sum does not individually converge.) Likewise, by (%
\protect\ref{eq:Realistic-C43N101Neg}) we see that the negative-%
$n$ terms in (\protect\ref{eq:Truncation-DeIdeal}) contribute an 
amount \setcounter{Ceqindent}{0}\protect\begin{eqnarray}\hspace{%
-1.3ex}&\displaystyle\mbox{$\protect\displaystyle\protect\frac{1%
}{a}$}\lim_{M\rightarrow\infty}\!\protect\left\{2\sum_{k=0}^M%
\mbox{$\protect\displaystyle\protect\frac{1}{202k+58}$}-\sum_{k=%
0}^{2M}\mbox{$\protect\displaystyle\protect\frac{1}{101k+58}$}%
\protect\right\}\!.\protect\nonumber\setlength{\Cscr}{\value{CEh%
t}\Ctenthex}\addtolength{\Cscr}{-1.0ex}\protect\raisebox{0ex}[%
\value{CEht}\Ctenthex][\Cscr]{}\protect\end{eqnarray}\setcounter
{CEht}{10}Putting these together, we then have \setcounter{Ceqin%
dent}{0}\protect\begin{eqnarray}\hspace{-1.3ex}&\displaystyle{}^%
{(1)\!}c_{43}^{101}=\mbox{$\protect\displaystyle\protect\frac{1}%
{a}$}\lim_{M\rightarrow\infty}\!\protect\left\{\,\sum_{k=0}^M\!%
\protect\left(\mbox{$\protect\displaystyle\protect\frac{2}{202k+%
43}$}+\mbox{$\protect\displaystyle\protect\frac{2}{202k+58}$}%
\protect\right)-\sum_{k=0}^{2M}\!\protect\left(\mbox{$\protect
\displaystyle\protect\frac{1}{101k+43}$}+\mbox{$\protect
\displaystyle\protect\frac{1}{101k+58}$}\protect\right)\!\protect
\right\}\!.\protect\nonumber\setlength{\Cscr}{\value{CEht}%
\Ctenthex}\addtolength{\Cscr}{-1.0ex}\protect\raisebox{0ex}[%
\value{CEht}\Ctenthex][\Cscr]{}\protect\end{eqnarray}\setcounter
{CEht}{10}Generalising to the case of general (odd)~$N$, and pos%
itive~$n$, we clearly obtain \setcounter{Ceqindent}{0}\protect
\begin{eqnarray}{}^{(1)\!}c_{\mbox{\scriptsize$n$ positive}}^{%
\mbox{\scriptsize$N\!$ odd}}\hspace{-1.3ex}&\displaystyle=&%
\hspace{-1.3ex}\mbox{$\protect\displaystyle\protect\frac{(-1)^{n%
+1}}{a}$}\!\lim_{M\rightarrow\infty}\!\setcounter{Cbscurr}{30}%
\setlength{\Cscr}{\value{Cbscurr}\Ctenthex}\addtolength{\Cscr}{-%
1.0ex}\protect\raisebox{0ex}[\value{Cbscurr}\Ctenthex][\Cscr]{}%
\hspace{-0ex}{\protect\left\{\setlength{\Cscr}{\value{Cbscurr}%
\Ctenthex}\addtolength{\Cscr}{-1.0ex}\protect\raisebox{0ex}[%
\value{Cbscurr}\Ctenthex][\Cscr]{}\protect\right.}\hspace{-0.25e%
x}\setlength{\Cscr}{\value{Cbscurr}\Ctenthex}\addtolength{\Cscr}%
{-1.0ex}\protect\raisebox{0ex}[\value{Cbscurr}\Ctenthex][\Cscr]{%
}\setcounter{CbsD}{\value{CbsC}}\setcounter{CbsC}{\value{CbsB}}%
\setcounter{CbsB}{\value{CbsA}}\setcounter{CbsA}{\value{Cbscurr}%
}\sum_{k=0}^M\!\protect\left(\mbox{$\protect\displaystyle\protect
\frac{2}{2Nk+n}$}+\mbox{$\protect\displaystyle\protect\frac{2}{2%
Nk+N-n}$}\protect\right)\setcounter{Ceqindent}{300}\protect
\nonumber\setlength{\Cscr}{\value{CEht}\Ctenthex}\addtolength{%
\Cscr}{-1.0ex}\protect\raisebox{0ex}[\value{CEht}\Ctenthex][\Cscr
]{}\\*[0ex]\protect\displaystyle\hspace{-1.3ex}&\displaystyle&%
\hspace{-1.3ex}{\protect\mbox{}}\hspace{\value{Ceqindent}%
\Ctenthex}-\sum_{k=0}^{2M}\!\protect\left(\mbox{$\protect
\displaystyle\protect\frac{1}{Nk+n}$}+\mbox{$\protect
\displaystyle\protect\frac{1}{Nk+N-n}$}\protect\right)\!\!%
\setlength{\Cscr}{\value{CbsA}\Ctenthex}\addtolength{\Cscr}{-1.0%
ex}\protect\raisebox{0ex}[\value{CbsA}\Ctenthex][\Cscr]{}\hspace
{-0.25ex}{\protect\left.\setlength{\Cscr}{\value{CbsA}\Ctenthex}%
\addtolength{\Cscr}{-1.0ex}\protect\raisebox{0ex}[\value{CbsA}%
\Ctenthex][\Cscr]{}\protect\right\}}\hspace{-0ex}\setlength{\Cscr
}{\value{CbsA}\Ctenthex}\addtolength{\Cscr}{-1.0ex}\protect
\raisebox{0ex}[\value{CbsA}\Ctenthex][\Cscr]{}\setcounter{CbsA}{%
\value{CbsB}}\setcounter{CbsB}{\value{CbsC}}\setcounter{CbsC}{%
\value{CbsD}}\setcounter{CbsD}{1}.\protect\nonumber\setlength{%
\Cscr}{\value{CEht}\Ctenthex}\addtolength{\Cscr}{-1.0ex}\protect
\raisebox{0ex}[\value{CEht}\Ctenthex][\Cscr]{}\protect\end{eqnar%
ray}\setcounter{CEht}{10}Again factorising $N$ from these denomi%
nators, we then obtain \setcounter{Ceqindent}{0}\protect\begin{e%
qnarray}{}^{(1)\!}c_{\mbox{\scriptsize$n$ positive}}^{\mbox{%
\scriptsize$N\!$ odd}}\hspace{-1.3ex}&\displaystyle=&\hspace{-1.%
3ex}\mbox{$\protect\displaystyle\protect\frac{(-1)^{n+1}}{Na}$}%
\!\lim_{M\rightarrow\infty}\!\setcounter{Cbscurr}{30}\setlength{%
\Cscr}{\value{Cbscurr}\Ctenthex}\addtolength{\Cscr}{-1.0ex}%
\protect\raisebox{0ex}[\value{Cbscurr}\Ctenthex][\Cscr]{}\hspace
{-0ex}{\protect\left\{\setlength{\Cscr}{\value{Cbscurr}\Ctenthex
}\addtolength{\Cscr}{-1.0ex}\protect\raisebox{0ex}[\value{Cbscur%
r}\Ctenthex][\Cscr]{}\protect\right.}\hspace{-0.25ex}\setlength{%
\Cscr}{\value{Cbscurr}\Ctenthex}\addtolength{\Cscr}{-1.0ex}%
\protect\raisebox{0ex}[\value{Cbscurr}\Ctenthex][\Cscr]{}%
\setcounter{CbsD}{\value{CbsC}}\setcounter{CbsC}{\value{CbsB}}%
\setcounter{CbsB}{\value{CbsA}}\setcounter{CbsA}{\value{Cbscurr}%
}\sum_{k=0}^M\!\protect\left(\mbox{$\protect\displaystyle\protect
\frac{1}{k+n/2N}$}+\mbox{$\protect\displaystyle\protect\frac{1}{%
k+1/2-n/2N}$}\protect\right)\setcounter{Ceqindent}{250}\protect
\nonumber\setlength{\Cscr}{\value{CEht}\Ctenthex}\addtolength{%
\Cscr}{-1.0ex}\protect\raisebox{0ex}[\value{CEht}\Ctenthex][\Cscr
]{}\\*[0ex]\protect\displaystyle\hspace{-1.3ex}&\displaystyle&%
\hspace{-1.3ex}{\protect\mbox{}}\hspace{\value{Ceqindent}%
\Ctenthex}-\sum_{k=0}^{2M}\!\protect\left(\mbox{$\protect
\displaystyle\protect\frac{1}{k+n/N}$}+\mbox{$\protect
\displaystyle\protect\frac{1}{k+1-n/N}$}\protect\right)\!\!%
\setlength{\Cscr}{\value{CbsA}\Ctenthex}\addtolength{\Cscr}{-1.0%
ex}\protect\raisebox{0ex}[\value{CbsA}\Ctenthex][\Cscr]{}\hspace
{-0.25ex}{\protect\left.\setlength{\Cscr}{\value{CbsA}\Ctenthex}%
\addtolength{\Cscr}{-1.0ex}\protect\raisebox{0ex}[\value{CbsA}%
\Ctenthex][\Cscr]{}\protect\right\}}\hspace{-0ex}\setlength{\Cscr
}{\value{CbsA}\Ctenthex}\addtolength{\Cscr}{-1.0ex}\protect
\raisebox{0ex}[\value{CbsA}\Ctenthex][\Cscr]{}\setcounter{CbsA}{%
\value{CbsB}}\setcounter{CbsB}{\value{CbsC}}\setcounter{CbsC}{%
\value{CbsD}}\setcounter{CbsD}{1}.\protect\nonumber\setlength{%
\Cscr}{\value{CEht}\Ctenthex}\addtolength{\Cscr}{-1.0ex}\protect
\raisebox{0ex}[\value{CEht}\Ctenthex][\Cscr]{}\protect\end{eqnar%
ray}\setcounter{CEht}{10}Fortunately, the mathematicians have al%
so evaluated this combination of infinite limits of sums for us, 
and again it is a remarkably simple result: \setcounter{Ceqinden%
t}{0}\protect\begin{eqnarray}\protect\left.\protect\begin{array}%
{rcl}\protect\displaystyle\hspace{-1.3ex}&&\hspace{-1.3ex}%
\protect\displaystyle\lim_{M\rightarrow\infty}\!\protect\left\{%
\,\sum_{k=0}^M\!\protect\left(\mbox{$\protect\displaystyle
\protect\frac{1}{k+n/2N}$}+\mbox{$\protect\displaystyle\protect
\frac{1}{k+1/2-n/2N}$}\protect\right)-\sum_{k=0}^{2M}\!\protect
\left(\mbox{$\protect\displaystyle\protect\frac{1}{k+n/N}$}+\mbox
{$\protect\displaystyle\protect\frac{1}{k+1-n/N}$}\protect\right
)\!\protect\right\}\setcounter{Ceqindent}{680}\setlength{\Cscr}{%
\value{CEht}\Ctenthex}\addtolength{\Cscr}{-1.0ex}\protect
\raisebox{0ex}[\value{CEht}\Ctenthex][\Cscr]{}\\*[0.55ex]\protect
\displaystyle\hspace{-1.3ex}&\protect\displaystyle&\hspace{-1.3e%
x}\protect\displaystyle{\protect\mbox{}}\hspace{\value{Ceqindent%
}\Ctenthex}=\setcounter{CEht}{40}\pi\csc\!\protect\left(\mbox{$%
\protect\displaystyle\protect\frac{\pi n}{N}$}\protect\right)\!.%
\vspace{-3ex}\setlength{\Cscr}{\value{CEht}\Ctenthex}\addtolength
{\Cscr}{-1.0ex}\protect\raisebox{0ex}[\value{CEht}\Ctenthex][%
\Cscr]{}\protect\end{array}\protect\right.\protect\label{eq:Exac%
t-InfiniteSumLinears2}\protect\end{eqnarray}\setcounter{CEht}{10%
}We thus find that \setcounter{Ceqindent}{0}\protect\begin{eqnar%
ray}\hspace{-1.3ex}&\displaystyle{}^{(1)\!}c_{\mbox{\scriptsize$%
n$ positive}}^{\mbox{\scriptsize$N\!$ odd}}=(-1)^{n+1}\mbox{$%
\protect\displaystyle\protect\frac{\pi}{Na}$}\csc\!\protect\left
(\mbox{$\protect\displaystyle\protect\frac{\pi n}{N}$}\protect
\right)\!.\protect\nonumber\setlength{\Cscr}{\value{CEht}%
\Ctenthex}\addtolength{\Cscr}{-1.0ex}\protect\raisebox{0ex}[%
\value{CEht}\Ctenthex][\Cscr]{}\protect\end{eqnarray}\setcounter
{CEht}{10}As before, this expression is odd under \mbox{$\protect
\displaystyle n\rightarrow-n$}, in accordance with (\protect\ref
{eq:Truncation-DeIdeal}), and so for a general position $n$ (for 
odd~$N$) we have the result \setcounter{Ceqindent}{0}\protect
\begin{eqnarray}\protect\left.\protect\begin{array}{rcl}\protect
\displaystyle\hspace{-1.3ex}&\protect\displaystyle{}^{(1)\!}c_{n%
\neq0}^{\mbox{\scriptsize$N\!$ odd}}=(-1)^{n+1}\mbox{$\protect
\displaystyle\protect\frac{\pi}{Na}$}\csc\!\protect\left(\mbox{$%
\protect\displaystyle\protect\frac{\pi n}{N}$}\protect\right)\!.%
\setlength{\Cscr}{\value{CEht}\Ctenthex}\addtolength{\Cscr}{-1.0%
ex}\protect\raisebox{0ex}[\value{CEht}\Ctenthex][\Cscr]{}\protect
\end{array}\protect\right.\protect\label{eq:Exact-C1OddN}\protect
\end{eqnarray}\setcounter{CEht}{10}Again, since \mbox{$\protect
\displaystyle\csc x\equiv1/\sin x\sim1/x$} for small~$x$, we fin%
d that far from the application of periodic boundary conditions 
we regain the infinite-lattice result: \setcounter{Ceqindent}{0}%
\protect\begin{eqnarray}\hspace{-1.3ex}&\displaystyle{}^{(1)\!}c%
_{n\neq0}^{\mbox{\scriptsize$N\!$ odd}}\rightarrow\mbox{$\protect
\displaystyle\protect\frac{(-1)^{n+1}}{na}$}\hspace{3ex}\mbox{as%
}\hspace{3ex}\mbox{$\protect\displaystyle\protect\frac{n}{N}$}%
\rightarrow0.\protect\nonumber\setlength{\Cscr}{\value{CEht}%
\Ctenthex}\addtolength{\Cscr}{-1.0ex}\protect\raisebox{0ex}[%
\value{CEht}\Ctenthex][\Cscr]{}\protect\end{eqnarray}\setcounter
{CEht}{10}However, in contrast to the case of even~$N$, for odd 
$N$ we find that, near the edge of the application of periodic b%
oundary conditions, namely, \mbox{$\protect\displaystyle n%
\rightarrow N/2$}, the absolute value of \mbox{$\protect
\displaystyle{}^{(1)\!}c_{n\neq0}^{\mbox{\scriptsize$N\!$ odd}}$%
} approaches the finite value \mbox{$\protect\displaystyle\pi/Na%
$}. If it weren't for the periodic boundary conditions, the abso%
lute value of the coefficient function at the limit of this boun%
dary (between lattice sites) would only have been \mbox{$\protect
\displaystyle1/na=2/Na$}; thus, the applications of periodic bou%
ndary conditions for odd $N$ has actually \mbox{}\protect\/{%
\protect\em increased\protect\/} the magnitude of the coefficien%
t function at the boundary by a factor of $\pi/2$, or around a 5%
7\% increase. Indeed, this is a general phenomenon: \setcounter{%
Ceqindent}{0}\protect\begin{eqnarray}\hspace{-1.3ex}&%
\displaystyle\protect\left|{}^{(1)\!}c_{n\neq0}^{\mbox{%
\scriptsize$N\!$ odd}}\protect\right|>\mbox{$\protect
\displaystyle\protect\frac{1}{na}$},\protect\nonumber\setlength{%
\Cscr}{\value{CEht}\Ctenthex}\addtolength{\Cscr}{-1.0ex}\protect
\raisebox{0ex}[\value{CEht}\Ctenthex][\Cscr]{}\protect\end{eqnar%
ray}\setcounter{CEht}{10}whereas \setcounter{Ceqindent}{0}%
\protect\begin{eqnarray}\hspace{-1.3ex}&\displaystyle\protect
\left|{}^{(1)\!}c_{n\neq0}^{\mbox{\scriptsize$N\!$ even}}\protect
\right|<\mbox{$\protect\displaystyle\protect\frac{1}{na}$},%
\protect\nonumber\setlength{\Cscr}{\value{CEht}\Ctenthex}%
\addtolength{\Cscr}{-1.0ex}\protect\raisebox{0ex}[\value{CEht}%
\Ctenthex][\Cscr]{}\protect\end{eqnarray}\setcounter{CEht}{10}al%
though in each case the magnitude of the inequality is relativel%
y mild, as we shall see shortly in \mbox{Sec.~$\:\!\!$}\protect
\ref{sect:Stochastic}.\par Let us now turn to the second-derivat%
ive operator. The expressions (\protect\ref{eq:SecondEx-EvenPos}%
) and (\protect\ref{eq:SecondEx-EvenNeg}) clearly combine to giv%
e us \setcounter{Ceqindent}{0}\protect\begin{eqnarray}\hspace{-1%
.3ex}&\displaystyle{}^{(2)\!}c_{43}^{100}=\mbox{$\protect
\displaystyle\protect\frac{2}{a^{2\!}}$}\sum_{k=0}^\infty\!%
\protect\left\{\mbox{$\protect\displaystyle\protect\frac{1}{(100%
k+43)^2}$}+\mbox{$\protect\displaystyle\protect\frac{1}{(100k+57%
)^2}$}\!\protect\right\}\!.\protect\nonumber\setlength{\Cscr}{%
\value{CEht}\Ctenthex}\addtolength{\Cscr}{-1.0ex}\protect
\raisebox{0ex}[\value{CEht}\Ctenthex][\Cscr]{}\protect\end{eqnar%
ray}\setcounter{CEht}{10}Going to general (even) $N$ and general 
\mbox{$\protect\displaystyle n\neq0$} (positive or negative, sin%
ce they come in with the same sign for the second-derivative), w%
e clearly have \setcounter{Ceqindent}{0}\protect\begin{eqnarray}%
\hspace{-1.3ex}&\displaystyle{}^{(2)\!}c_{n\neq0}^{\mbox{%
\scriptsize$N\!$ even}}=\mbox{$\protect\displaystyle\protect\frac
{2(-1)^{n+1}}{a^{2\!}}$}\!\sum_{k=0}^\infty\!\protect\left\{\mbox
{$\protect\displaystyle\protect\frac{1}{(Nk+n)^2}$}+\mbox{$%
\protect\displaystyle\protect\frac{1}{(Nk+N-n)^2}$}\!\protect
\right\}\!,\protect\nonumber\setlength{\Cscr}{\value{CEht}%
\Ctenthex}\addtolength{\Cscr}{-1.0ex}\protect\raisebox{0ex}[%
\value{CEht}\Ctenthex][\Cscr]{}\protect\end{eqnarray}\setcounter
{CEht}{10}and factorising out $N^2$ from the denominators we hav%
e \setcounter{Ceqindent}{0}\protect\begin{eqnarray}\hspace{-1.3e%
x}&\displaystyle{}^{(2)\!}c_{n\neq0}^{\mbox{\scriptsize$N\!$ eve%
n}}=\mbox{$\protect\displaystyle\protect\frac{2(-1)^{n+1}}{(Na)^%
{2\!}}$}\!\sum_{k=0}^\infty\!\protect\left\{\mbox{$\protect
\displaystyle\protect\frac{1}{(k+n/N)^2}$}+\mbox{$\protect
\displaystyle\protect\frac{1}{(k+1-n/N)^2}$}\!\protect\right\}\!%
.\protect\nonumber\setlength{\Cscr}{\value{CEht}\Ctenthex}%
\addtolength{\Cscr}{-1.0ex}\protect\raisebox{0ex}[\value{CEht}%
\Ctenthex][\Cscr]{}\protect\end{eqnarray}\setcounter{CEht}{10}Ag%
ain, the mathematicians give us a magic result for the infinite 
sum: \setcounter{Ceqindent}{0}\protect\begin{eqnarray}\protect
\left.\protect\begin{array}{rcl}\protect\displaystyle\hspace{-1.%
3ex}&\protect\displaystyle\sum_{k=0}^\infty\!\protect\left\{\mbox
{$\protect\displaystyle\protect\frac{1}{(k+n/N)^2}$}+\mbox{$%
\protect\displaystyle\protect\frac{1}{(k+1-n/N)^2}$}\!\protect
\right\}={\pi}{}^{\raisebox{-0.25ex}{$\scriptstyle2\:\!$}}{\csc}%
{}^{\raisebox{-0.25ex}{$\scriptstyle2\!$}}\!\protect\left(\mbox{%
$\protect\displaystyle\protect\frac{\pi n}{N}$}\protect\right)\!%
,\setlength{\Cscr}{\value{CEht}\Ctenthex}\addtolength{\Cscr}{-1.%
0ex}\protect\raisebox{0ex}[\value{CEht}\Ctenthex][\Cscr]{}%
\protect\end{array}\protect\right.\protect\label{eq:Exact-Infini%
teSumSquared1}\protect\end{eqnarray}\setcounter{CEht}{10}so that 
\setcounter{Ceqindent}{0}\protect\begin{eqnarray}\protect\left.%
\protect\begin{array}{rcl}\protect\displaystyle\hspace{-1.3ex}&%
\protect\displaystyle{}^{(2)\!}c_{n\neq0}^{\mbox{\scriptsize$N\!%
$ even}}=2(-1)^{n+1\:\!\!}\!\protect\left(\mbox{$\protect
\displaystyle\protect\frac{\pi}{Na}$}\protect\right)^{\!\!2}\!{%
\csc}{}^{\raisebox{-0.25ex}{$\scriptstyle2\!$}}\!\protect\left(%
\mbox{$\protect\displaystyle\protect\frac{\pi n}{N}$}\protect
\right)\!.\setlength{\Cscr}{\value{CEht}\Ctenthex}\addtolength{%
\Cscr}{-1.0ex}\protect\raisebox{0ex}[\value{CEht}\Ctenthex][\Cscr
]{}\protect\end{array}\protect\right.\protect\label{eq:Exact-C2E%
venN}\protect\end{eqnarray}\setcounter{CEht}{10}{}From\ the way 
we have constructed it, this expression clearly also applies to 
the (single) position \mbox{$\protect\displaystyle n=+N/2$}, tha%
t is half a lattice away from the position \mbox{$\protect
\displaystyle n=0$} at which we are applying the derivative. Cle%
arly, far from the application of periodic boundary conditions, 
we again regain from (\protect\ref{eq:Exact-C2EvenN}) the infini%
te-lattice result (\protect\ref{eq:Truncation-De2Ideal}).\par Fo%
r \mbox{$\protect\displaystyle n=0$}, the expression (\protect
\ref{eq:SecondEx-EvenZero}) can be written \setcounter{Ceqindent%
}{0}\protect\begin{eqnarray}\hspace{-1.3ex}&\displaystyle{}^{(2)%
\!}c_0^{\mbox{\scriptsize$N\!$ even}}=-\mbox{$\protect
\displaystyle\protect\frac{2}{a^{2\!}}$}\!\protect\left\{\mbox{$%
\protect\displaystyle\protect\frac{{\pi}{}^{\raisebox{-0.25ex}{$%
\scriptstyle2\!$}}}{6}$}+\mbox{$\protect\displaystyle\protect
\frac{2}{N^{2\!\!}}$}\sum_{k=1}^\infty\mbox{$\protect
\displaystyle\protect\frac{1}{k^{2\!\!}}$}^{\,}\protect\right\}%
\!,\protect\nonumber\setlength{\Cscr}{\value{CEht}\Ctenthex}%
\addtolength{\Cscr}{-1.0ex}\protect\raisebox{0ex}[\value{CEht}%
\Ctenthex][\Cscr]{}\protect\end{eqnarray}\setcounter{CEht}{10}an%
d since \setcounter{Ceqindent}{0}\protect\begin{eqnarray}\hspace
{-1.3ex}&\displaystyle\sum_{k=1}^\infty\mbox{$\protect
\displaystyle\protect\frac{1}{k^{2\!\!}}$}\equiv\zeta(2)=\mbox{$%
\protect\displaystyle\protect\frac{{\pi}{}^{\raisebox{-0.25ex}{$%
\scriptstyle2\!\!$}}}{6}$},\protect\nonumber\setlength{\Cscr}{%
\value{CEht}\Ctenthex}\addtolength{\Cscr}{-1.0ex}\protect
\raisebox{0ex}[\value{CEht}\Ctenthex][\Cscr]{}\protect\end{eqnar%
ray}\setcounter{CEht}{10}we find \setcounter{Ceqindent}{0}%
\protect\begin{eqnarray}\protect\left.\protect\begin{array}{rcl}%
\protect\displaystyle\hspace{-1.3ex}&\protect\displaystyle{}^{(2%
)\!}c_0^{\mbox{\scriptsize$N\!$ even}}=-\mbox{$\protect
\displaystyle\protect\frac{{\pi}{}^{\raisebox{-0.25ex}{$%
\scriptstyle2$}}}{3a^{2\!}}$}\!\protect\left(1+\mbox{$\protect
\displaystyle\protect\frac{2}{N^{2\!\!}}$}\protect\right)\!.%
\setlength{\Cscr}{\value{CEht}\Ctenthex}\addtolength{\Cscr}{-1.0%
ex}\protect\raisebox{0ex}[\value{CEht}\Ctenthex][\Cscr]{}\protect
\end{array}\protect\right.\protect\label{eq:Exact-C20EvenN}%
\protect\end{eqnarray}\setcounter{CEht}{10}It can be shown that 
the additional contribution here for finite $N$ is equal to the 
difference between the finite-lattice result (\protect\ref{eq:Ex%
act-C2EvenN}) and the infinite-lattice result (\protect\ref{eq:T%
runcation-De2Ideal}) for \mbox{$\protect\displaystyle n\neq0$} i%
f we assume the expression to hold true for real $n$ and take th%
e limit \mbox{$\protect\displaystyle n\rightarrow0$}.\par Finall%
y, we consider the case of odd $N$. The sum in (\protect\ref{eq:%
SecondEx-OddPos}) can be written \setcounter{Ceqindent}{0}%
\protect\begin{eqnarray}\hspace{-1.3ex}&\displaystyle\mbox{$%
\protect\displaystyle\protect\frac{2}{a^{2\!\!}}$}\protect\left
\{2\sum_{k=0}^\infty\mbox{$\protect\displaystyle\protect\frac{1}%
{(202k+43)^2}$}-\sum_{k=0}^\infty\mbox{$\protect\displaystyle
\protect\frac{1}{(101k+43)^2}$}\protect\right\}\!,\protect
\nonumber\setlength{\Cscr}{\value{CEht}\Ctenthex}\addtolength{%
\Cscr}{-1.0ex}\protect\raisebox{0ex}[\value{CEht}\Ctenthex][\Cscr
]{}\protect\end{eqnarray}\setcounter{CEht}{10}where in this case 
we don't need to be careful about the limiting process because e%
ach sum is separately convergent. Likewise, the sum in (\protect
\ref{eq:SecondEx-OddNeg}) can be written \setcounter{Ceqindent}{%
0}\protect\begin{eqnarray}\hspace{-1.3ex}&\displaystyle-\mbox{$%
\protect\displaystyle\protect\frac{2}{a^{2\!\!}}$}\protect\left
\{2\sum_{k=0}^\infty\mbox{$\protect\displaystyle\protect\frac{1}%
{(202k+58)^2}$}-\sum_{k=0}^\infty\mbox{$\protect\displaystyle
\protect\frac{1}{(101k+58)^2}$}\protect\right\}\!,\protect
\nonumber\setlength{\Cscr}{\value{CEht}\Ctenthex}\addtolength{%
\Cscr}{-1.0ex}\protect\raisebox{0ex}[\value{CEht}\Ctenthex][\Cscr
]{}\protect\end{eqnarray}\setcounter{CEht}{10}Combining these ex%
pressions, we then have \setcounter{Ceqindent}{0}\protect\begin{%
eqnarray}\hspace{-1.3ex}&\displaystyle{}^{(2)\!}c_{43}^{101}=%
\mbox{$\protect\displaystyle\protect\frac{2}{a^{2\!\!}}$}\sum_{k%
=0}^\infty\!\protect\left\{\mbox{$\protect\displaystyle\protect
\frac{2}{(202k+43)^2}$}-\mbox{$\protect\displaystyle\protect\frac
{2}{(202k+58)^2}$}-\mbox{$\protect\displaystyle\protect\frac{1}{%
(101k+43)^2}$}+\mbox{$\protect\displaystyle\protect\frac{1}{(101%
k+58)^2}$}\protect\right\}\!.\protect\nonumber\setlength{\Cscr}{%
\value{CEht}\Ctenthex}\addtolength{\Cscr}{-1.0ex}\protect
\raisebox{0ex}[\value{CEht}\Ctenthex][\Cscr]{}\protect\end{eqnar%
ray}\setcounter{CEht}{10}For general (odd) $N$ and \mbox{$%
\protect\displaystyle n\neq0$}, this becomes \setcounter{Ceqinde%
nt}{0}\protect\begin{eqnarray}{}^{(2)\!}c_{n\neq0}^{\mbox{%
\scriptsize$N\!$ odd}}\hspace{-1.3ex}&\displaystyle=&\hspace{-1.%
3ex}\mbox{$\protect\displaystyle\protect\frac{2(-1)^{n+1}}{a^{2%
\!\!}}$}\!\sum_{k=0}^\infty\!\setcounter{Cbscurr}{30}\setlength{%
\Cscr}{\value{Cbscurr}\Ctenthex}\addtolength{\Cscr}{-1.0ex}%
\protect\raisebox{0ex}[\value{Cbscurr}\Ctenthex][\Cscr]{}\hspace
{-0ex}{\protect\left\{\setlength{\Cscr}{\value{Cbscurr}\Ctenthex
}\addtolength{\Cscr}{-1.0ex}\protect\raisebox{0ex}[\value{Cbscur%
r}\Ctenthex][\Cscr]{}\protect\right.}\hspace{-0.25ex}\setlength{%
\Cscr}{\value{Cbscurr}\Ctenthex}\addtolength{\Cscr}{-1.0ex}%
\protect\raisebox{0ex}[\value{Cbscurr}\Ctenthex][\Cscr]{}%
\setcounter{CbsD}{\value{CbsC}}\setcounter{CbsC}{\value{CbsB}}%
\setcounter{CbsB}{\value{CbsA}}\setcounter{CbsA}{\value{Cbscurr}%
}\mbox{$\protect\displaystyle\protect\frac{2}{(2Nk+n)^2}$}-\mbox
{$\protect\displaystyle\protect\frac{2}{(2Nk+N-n)^2}$}\setcounter
{Ceqindent}{250}\protect\nonumber\setlength{\Cscr}{\value{CEht}%
\Ctenthex}\addtolength{\Cscr}{-1.0ex}\protect\raisebox{0ex}[%
\value{CEht}\Ctenthex][\Cscr]{}\\*[0ex]\protect\displaystyle
\hspace{-1.3ex}&\displaystyle&\hspace{-1.3ex}{\protect\mbox{}}%
\hspace{\value{Ceqindent}\Ctenthex}-\mbox{$\protect\displaystyle
\protect\frac{1}{(Nk+n)^2}$}+\mbox{$\protect\displaystyle\protect
\frac{1}{(Nk+N-n)^2}$}\setlength{\Cscr}{\value{CbsA}\Ctenthex}%
\addtolength{\Cscr}{-1.0ex}\protect\raisebox{0ex}[\value{CbsA}%
\Ctenthex][\Cscr]{}\hspace{-0.25ex}{\protect\left.\setlength{%
\Cscr}{\value{CbsA}\Ctenthex}\addtolength{\Cscr}{-1.0ex}\protect
\raisebox{0ex}[\value{CbsA}\Ctenthex][\Cscr]{}\protect\right\}}%
\hspace{-0ex}\setlength{\Cscr}{\value{CbsA}\Ctenthex}\addtolength
{\Cscr}{-1.0ex}\protect\raisebox{0ex}[\value{CbsA}\Ctenthex][%
\Cscr]{}\setcounter{CbsA}{\value{CbsB}}\setcounter{CbsB}{\value{%
CbsC}}\setcounter{CbsC}{\value{CbsD}}\setcounter{CbsD}{1},%
\protect\nonumber\setlength{\Cscr}{\value{CEht}\Ctenthex}%
\addtolength{\Cscr}{-1.0ex}\protect\raisebox{0ex}[\value{CEht}%
\Ctenthex][\Cscr]{}\protect\end{eqnarray}\setcounter{CEht}{10}an%
d on factorising out $N^2$ we get \setcounter{Ceqindent}{0}%
\protect\begin{eqnarray}{}^{(2)\!}c_{n\neq0}^{\mbox{\scriptsize$%
N\!$ odd}}\hspace{-1.3ex}&\displaystyle=&\hspace{-1.3ex}\mbox{$%
\protect\displaystyle\protect\frac{2(-1)^{n+1}}{(Na)^{2\!\!}}$}%
\!\sum_{k=0}^\infty\!\setcounter{Cbscurr}{30}\setlength{\Cscr}{%
\value{Cbscurr}\Ctenthex}\addtolength{\Cscr}{-1.0ex}\protect
\raisebox{0ex}[\value{Cbscurr}\Ctenthex][\Cscr]{}\hspace{-0ex}{%
\protect\left\{\setlength{\Cscr}{\value{Cbscurr}\Ctenthex}%
\addtolength{\Cscr}{-1.0ex}\protect\raisebox{0ex}[\value{Cbscurr%
}\Ctenthex][\Cscr]{}\protect\right.}\hspace{-0.25ex}\setlength{%
\Cscr}{\value{Cbscurr}\Ctenthex}\addtolength{\Cscr}{-1.0ex}%
\protect\raisebox{0ex}[\value{Cbscurr}\Ctenthex][\Cscr]{}%
\setcounter{CbsD}{\value{CbsC}}\setcounter{CbsC}{\value{CbsB}}%
\setcounter{CbsB}{\value{CbsA}}\setcounter{CbsA}{\value{Cbscurr}%
}\mbox{$\protect\displaystyle\protect\frac{1/2}{(k+n/2N)^2}$}-%
\mbox{$\protect\displaystyle\protect\frac{1/2}{(k+1/2-n/2N)^2}$}%
\setcounter{Ceqindent}{250}\protect\nonumber\setlength{\Cscr}{%
\value{CEht}\Ctenthex}\addtolength{\Cscr}{-1.0ex}\protect
\raisebox{0ex}[\value{CEht}\Ctenthex][\Cscr]{}\\*[0ex]\protect
\displaystyle\hspace{-1.3ex}&\displaystyle&\hspace{-1.3ex}{%
\protect\mbox{}}\hspace{\value{Ceqindent}\Ctenthex}-\mbox{$%
\protect\displaystyle\protect\frac{1}{(k+n/N)^2}$}+\mbox{$%
\protect\displaystyle\protect\frac{1}{(k+1-n/N)^2}$}\setlength{%
\Cscr}{\value{CbsA}\Ctenthex}\addtolength{\Cscr}{-1.0ex}\protect
\raisebox{0ex}[\value{CbsA}\Ctenthex][\Cscr]{}\hspace{-0.25ex}{%
\protect\left.\setlength{\Cscr}{\value{CbsA}\Ctenthex}%
\addtolength{\Cscr}{-1.0ex}\protect\raisebox{0ex}[\value{CbsA}%
\Ctenthex][\Cscr]{}\protect\right\}}\hspace{-0ex}\setlength{\Cscr
}{\value{CbsA}\Ctenthex}\addtolength{\Cscr}{-1.0ex}\protect
\raisebox{0ex}[\value{CbsA}\Ctenthex][\Cscr]{}\setcounter{CbsA}{%
\value{CbsB}}\setcounter{CbsB}{\value{CbsC}}\setcounter{CbsC}{%
\value{CbsD}}\setcounter{CbsD}{1}.\protect\nonumber\setlength{%
\Cscr}{\value{CEht}\Ctenthex}\addtolength{\Cscr}{-1.0ex}\protect
\raisebox{0ex}[\value{CEht}\Ctenthex][\Cscr]{}\protect\end{eqnar%
ray}\setcounter{CEht}{10}Again, the mathematicians have evaluate%
d this sum for us: \setcounter{Ceqindent}{0}\protect\begin{eqnar%
ray}\protect\left.\protect\begin{array}{rcl}\protect\displaystyle
\hspace{-1.3ex}&&\hspace{-1.3ex}\protect\displaystyle\sum_{k=0}^%
\infty\!\setcounter{Cbscurr}{30}\setlength{\Cscr}{\value{Cbscurr%
}\Ctenthex}\addtolength{\Cscr}{-1.0ex}\protect\raisebox{0ex}[%
\value{Cbscurr}\Ctenthex][\Cscr]{}\hspace{-0ex}{\protect\left\{%
\setlength{\Cscr}{\value{Cbscurr}\Ctenthex}\addtolength{\Cscr}{-%
1.0ex}\protect\raisebox{0ex}[\value{Cbscurr}\Ctenthex][\Cscr]{}%
\protect\right.}\hspace{-0.25ex}\setlength{\Cscr}{\value{Cbscurr%
}\Ctenthex}\addtolength{\Cscr}{-1.0ex}\protect\raisebox{0ex}[%
\value{Cbscurr}\Ctenthex][\Cscr]{}\setcounter{CbsD}{\value{CbsC}%
}\setcounter{CbsC}{\value{CbsB}}\setcounter{CbsB}{\value{CbsA}}%
\setcounter{CbsA}{\value{Cbscurr}}\mbox{$\protect\displaystyle
\protect\frac{1/2}{(k+n/2N)^2}$}-\mbox{$\protect\displaystyle
\protect\frac{1/2}{(k+1/2-n/2N)^2}$}\setcounter{Ceqindent}{200}%
\setlength{\Cscr}{\value{CEht}\Ctenthex}\addtolength{\Cscr}{-1.0%
ex}\protect\raisebox{0ex}[\value{CEht}\Ctenthex][\Cscr]{}\\*[0.5%
5ex]\protect\displaystyle\hspace{-1.3ex}&\protect\displaystyle&%
\hspace{-1.3ex}\protect\displaystyle{\protect\mbox{}}\hspace{%
\value{Ceqindent}\Ctenthex}-\mbox{$\protect\displaystyle\protect
\frac{1}{(k+n/N)^2}$}+\mbox{$\protect\displaystyle\protect\frac{%
1}{(k+1-n/N)^2}$}\setlength{\Cscr}{\value{CbsA}\Ctenthex}%
\addtolength{\Cscr}{-1.0ex}\protect\raisebox{0ex}[\value{CbsA}%
\Ctenthex][\Cscr]{}\hspace{-0.25ex}{\protect\left.\setlength{%
\Cscr}{\value{CbsA}\Ctenthex}\addtolength{\Cscr}{-1.0ex}\protect
\raisebox{0ex}[\value{CbsA}\Ctenthex][\Cscr]{}\protect\right\}}%
\hspace{-0ex}\setlength{\Cscr}{\value{CbsA}\Ctenthex}\addtolength
{\Cscr}{-1.0ex}\protect\raisebox{0ex}[\value{CbsA}\Ctenthex][%
\Cscr]{}\setcounter{CbsA}{\value{CbsB}}\setcounter{CbsB}{\value{%
CbsC}}\setcounter{CbsC}{\value{CbsD}}\setcounter{CbsD}{1}={\pi}{%
}^{\raisebox{-0.25ex}{$\scriptstyle2\:\!\!$}}\csc\!\protect\left
(\mbox{$\protect\displaystyle\protect\frac{\pi n}{N}$}\protect
\right)^{\!}\cot\!\protect\left(\mbox{$\protect\displaystyle
\protect\frac{\pi n}{N}$}\protect\right)\!,\setlength{\Cscr}{%
\value{CEht}\Ctenthex}\addtolength{\Cscr}{-1.0ex}\protect
\raisebox{0ex}[\value{CEht}\Ctenthex][\Cscr]{}\protect\end{array%
}\protect\right.\protect\label{eq:Exact-InfiniteSumSquared2}%
\protect\end{eqnarray}\setcounter{CEht}{10}so that \setcounter{C%
eqindent}{0}\protect\begin{eqnarray}\protect\left.\protect\begin
{array}{rcl}\protect\displaystyle\hspace{-1.3ex}&\protect
\displaystyle{}^{(2)\!}c_{n\neq0}^{\mbox{\scriptsize$N\!$ odd}}=%
2(-1)^{n+1\:\!\!}\!\protect\left(\mbox{$\protect\displaystyle
\protect\frac{\pi}{Na}$}\protect\right)^{\!\!2}\!\csc\!\protect
\left(\mbox{$\protect\displaystyle\protect\frac{\pi n}{N}$}%
\protect\right)^{\!}\cot\!\protect\left(\mbox{$\protect
\displaystyle\protect\frac{\pi n}{N}$}\protect\right)\!.%
\setlength{\Cscr}{\value{CEht}\Ctenthex}\addtolength{\Cscr}{-1.0%
ex}\protect\raisebox{0ex}[\value{CEht}\Ctenthex][\Cscr]{}\protect
\end{array}\protect\right.\protect\label{eq:Exact-C2OddN}\protect
\end{eqnarray}\setcounter{CEht}{10}For \mbox{$\protect
\displaystyle n=0$}, the expression (\protect\ref{eq:SecondEx-Od%
dZero}) can be written \setcounter{Ceqindent}{0}\protect\begin{e%
qnarray}\hspace{-1.3ex}&\displaystyle{}^{(2)\!}c_0^{\mbox{%
\scriptsize$N\!$ odd}}=-\mbox{$\protect\displaystyle\protect\frac
{2}{a^{2\!}}$}\!\protect\left\{\mbox{$\protect\displaystyle
\protect\frac{{\pi}{}^{\raisebox{-0.25ex}{$\scriptstyle2\!$}}}{6%
}$}-\mbox{$\protect\displaystyle\protect\frac{2}{N^{2\!\!}}$}\sum
_{k=1}^\infty{}^{\!}\mbox{$\protect\displaystyle\protect\frac{(-%
1)^{k+1}}{k^2}$}^{\,}\protect\right\}\!,\protect\nonumber
\setlength{\Cscr}{\value{CEht}\Ctenthex}\addtolength{\Cscr}{-1.0%
ex}\protect\raisebox{0ex}[\value{CEht}\Ctenthex][\Cscr]{}\protect
\end{eqnarray}\setcounter{CEht}{10}and so, using (\protect\ref{e%
q:Periodic-Eta2}), we find \setcounter{Ceqindent}{0}\protect
\begin{eqnarray}\protect\left.\protect\begin{array}{rcl}\protect
\displaystyle\hspace{-1.3ex}&\protect\displaystyle{}^{(2)\!}c_0^%
{\mbox{\scriptsize$N\!$ odd}}=-\mbox{$\protect\displaystyle
\protect\frac{{\pi}{}^{\raisebox{-0.25ex}{$\scriptstyle2$}}}{3a^%
{2\!}}$}\!\protect\left(1-\mbox{$\protect\displaystyle\protect
\frac{1}{N^{2\!\!}}$}\protect\right)\!.\setlength{\Cscr}{\value{%
CEht}\Ctenthex}\addtolength{\Cscr}{-1.0ex}\protect\raisebox{0ex}%
[\value{CEht}\Ctenthex][\Cscr]{}\protect\end{array}\protect\right
.\protect\label{eq:Exact-C20OddN}\protect\end{eqnarray}%
\setcounter{CEht}{10}The expressions (\protect\ref{eq:Exact-C2Od%
dN}) and (\protect\ref{eq:Exact-C20OddN}) are in agreement with 
those listed explicitly by \protect\ref{au:Drell1976}~[\ref{cit:%
Drell1976}] in the first of their seminal papers. Note, however, 
that these authors only considered the case of an odd number of 
lattice sites (which is written as \mbox{$\protect\displaystyle2%
N+1$} in their notation); and as no explicit expressions for the 
first-derivative operation in position space are provided in [%
\ref{cit:Drell1976}], further comparisons with the expressions l%
isted above are not possible.\par Clearly, the expressions above 
for the second-derivative operator on finite lattices also reduc%
e to the infinite-lattice results for \mbox{$\protect
\displaystyle n/N\rightarrow0$}, as was the case for the first-d%
erivative operators. Near the application of periodic boundary c%
onditions, the even-$N$ expression (\protect\ref{eq:Exact-C2Even%
N}) approaches a finite limit, whereas the odd-$N$ expression (%
\protect\ref{eq:Exact-C2OddN}) approaches zero, which is a rever%
sal of r\^oles from the case of the first-derivative operators (%
for which the even-$N$ coefficient vanished linearly, and the od%
d-$N$ coefficient approached a finite value).\par\refstepcounter
{section}\vspace{1.5\baselineskip}\par{\centering\bf\thesection. 
Stochastic implementation of the finite-lattice SLAC derivative 
operators\\*[0.5\baselineskip]}\protect\indent\label{sect:Stocha%
stic}Let us now turn to the reason why we are considering the pr%
actical implementation of the SLAC derivative operators at all, 
namely, my proposal in [\ref{cit:Costella2002}]\ that they be im%
plemented in a stochastic fashion. How must this proposal be mod%
ified to accommodate the corrections obtained in \mbox{Sec.~$\:%
\!\!$}\protect\ref{sect:Exact} to incorporate exactly the implem%
entation of periodic boundary conditions for a finite lattice?%
\par I propose the following: that the probabilistic weight assi%
gned to the computation of any difference (sum) in the first (se%
cond) SLAC derivative operator at distance $n$ be calculated usi%
ng the \mbox{}\protect\/{\protect\em same\protect\/} value $1/n$ 
($2/n^2$) as was proposed in [\ref{cit:Costella2002}]\ on the ba%
sis of the truncation of the infinite-lattice operators, but tha%
t the result of such sum (difference)---if calculated---be multi%
plied by a ``correction factor'' that incorporates exactly the e%
ffects of restricting the formalism to a finite lattice and appl%
ying periodic boundary conditions.\par My reasons for proposing 
this solution are twofold. Firstly, as noted in \mbox{Sec.~$\:\!%
\!$}\protect\ref{sect:Exact} (and as will become explicit below)%
, the required ``correction factors'' are, in all cases, simply 
numbers of order unity, which do not affect the fluctuation prop%
erties of the proposed stochastic implementation in any function%
ally critical fashion; we simply weight some positions a little 
more, or less, than for the case of an infinite lattice. Secondl%
y, it is clear that if this is all that we need to do, then it w%
ould be silly to tamper with the extraordinarily efficient metho%
d of random selection described in [\ref{cit:Costella2002}]\ for 
probabilistic weight functions of $1/n$ or $2/n^2$\mbox{$\!$}, b%
ecause the decision to compute or not compute needs to be made f%
or every distance $n$ for every application of the derivative op%
erator, and may, if blown out to a set of floating-point computa%
tions, become the time-critical part of the algorithm that would 
limit its practical implementation on a digital computer. (It is 
important to note that the floating-point multiplication of the 
result of the sum or difference by the correction factor need on%
ly be done \mbox{}\protect\/{\protect\em if the sum or differenc%
e is actually computed\protect\/}, and hence the average number 
of extra floating-point operations according to this proposed so%
lution is itself proportional to $\ln N$, rather than being prop%
ortional to $N$ if the probabilistic weights were instead modifi%
ed.)\par{}From\ the results obtained in \mbox{Sec.~$\:\!\!$}%
\protect\ref{sect:Exact}, it is simple to write down expressions 
for these correction factors. If we define \setcounter{Ceqindent%
}{0}\protect\begin{eqnarray}\hspace{-1.3ex}&\displaystyle{}^{(m)%
\!}\kappa_n^N\protect\nonumber\setlength{\Cscr}{\value{CEht}%
\Ctenthex}\addtolength{\Cscr}{-1.0ex}\protect\raisebox{0ex}[%
\value{CEht}\Ctenthex][\Cscr]{}\protect\end{eqnarray}\setcounter
{CEht}{10}to be the correction factor for the distance-$n$ sum o%
r difference for the $m$-th order SLAC derivative operator on a 
(one-dimensional) lattice of $N$ sites, then clearly \setcounter
{Ceqindent}{0}\protect\begin{eqnarray}\hspace{-1.3ex}&%
\displaystyle{}^{(m)\!}\kappa_n^N\equiv\mbox{$\protect
\displaystyle\protect\frac{{}^{(m)\!}c_n^N}{{}^{(m)\!}c_n^\infty
}$}.\protect\nonumber\setlength{\Cscr}{\value{CEht}\Ctenthex}%
\addtolength{\Cscr}{-1.0ex}\protect\raisebox{0ex}[\value{CEht}%
\Ctenthex][\Cscr]{}\protect\end{eqnarray}\setcounter{CEht}{10}{}%
 From\ the results (\protect\ref{eq:Exact-C1EvenN}), (\protect
\ref{eq:Exact-C1OddN}), (\protect\ref{eq:Exact-C2EvenN})\ and (%
\protect\ref{eq:Exact-C2OddN}) we then simply have \setcounter{C%
Elett}{0}\protect\refstepcounter{equation}\protect\label{eq:Stoc%
hastic-Final}\renewcommand\theequation{\arabic{equation}\alph{CE%
lett}}\setcounter{Ceqindent}{0}\protect\begin{eqnarray}{}^{(1)\!%
}\kappa_{n>0}^{\mbox{\scriptsize$N\!$ even}}\:\!\!\hspace{-1.3ex%
}&\displaystyle=&\hspace{-1.3ex}\alpha\cot\alpha,\protect
\stepcounter{CElett}\addtocounter{equation}{-1}\protect\label{eq%
:Stochastic-Ka1Even}\setlength{\Cscr}{\value{CEht}\Ctenthex}%
\addtolength{\Cscr}{-1.0ex}\protect\raisebox{0ex}[\value{CEht}%
\Ctenthex][\Cscr]{}\\*[0ex]\protect\displaystyle{}^{(1)\!}\kappa
_{n>0}^{\mbox{\scriptsize$N\!$ odd}}\:\!\!\hspace{-1.3ex}&%
\displaystyle=&\hspace{-1.3ex}\alpha\csc\alpha,\protect
\stepcounter{CElett}\addtocounter{equation}{-1}\protect\label{eq%
:Stochastic-Ka1Odd}\setlength{\Cscr}{\value{CEht}\Ctenthex}%
\addtolength{\Cscr}{-1.0ex}\protect\raisebox{0ex}[\value{CEht}%
\Ctenthex][\Cscr]{}\\*[0ex]\protect\displaystyle{}^{(2)\!}\kappa
_{n>0}^{\mbox{\scriptsize$N\!$ even}}\:\!\!\hspace{-1.3ex}&%
\displaystyle=&\hspace{-1.3ex}{\alpha}{}^{\raisebox{-0.25ex}{$%
\scriptstyle2\:\!$}}{\csc}{}^{\raisebox{-0.25ex}{$\scriptstyle2$%
}}\alpha,\protect\stepcounter{CElett}\addtocounter{equation}{-1}%
\protect\label{eq:Stochastic-Ka2Even}\setlength{\Cscr}{\value{CE%
ht}\Ctenthex}\addtolength{\Cscr}{-1.0ex}\protect\raisebox{0ex}[%
\value{CEht}\Ctenthex][\Cscr]{}\\*[0ex]\protect\displaystyle{}^{%
(2)\!}\kappa_{n>0}^{\mbox{\scriptsize$N\!$ odd}}\:\!\!\hspace{-1%
.3ex}&\displaystyle=&\hspace{-1.3ex}{\alpha}{}^{\raisebox{-0.25e%
x}{$\scriptstyle2\!$}}\csc\alpha\cot\alpha,\protect\stepcounter{%
CElett}\addtocounter{equation}{-1}\protect\label{eq:Stochastic-K%
a2Odd}\setlength{\Cscr}{\value{CEht}\Ctenthex}\addtolength{\Cscr
}{-1.0ex}\protect\raisebox{0ex}[\value{CEht}\Ctenthex][\Cscr]{}%
\protect\end{eqnarray}\setcounter{CEht}{10}\renewcommand
\theequation{\arabic{equation}}where \setcounter{Ceqindent}{0}%
\protect\begin{eqnarray}\protect\left.\protect\begin{array}{rcl}%
\protect\displaystyle\hspace{-1.3ex}&\protect\displaystyle\hspace
{6ex}\alpha\equiv\mbox{$\protect\displaystyle\protect\frac{\pi n%
}{N}$}\hspace{4ex}\protect\left(0<\alpha\leq\mbox{$\protect
\displaystyle\protect\frac{\pi}{2}$}\protect\right)\!,\setlength
{\Cscr}{\value{CEht}\Ctenthex}\addtolength{\Cscr}{-1.0ex}\protect
\raisebox{0ex}[\value{CEht}\Ctenthex][\Cscr]{}\protect\end{array%
}\protect\right.\protect\label{eq:Stochastic-Alpha}\protect\end{%
eqnarray}\setcounter{CEht}{10}and from (\protect\ref{eq:Exact-C2%
0EvenN}) and (\protect\ref{eq:Exact-C20OddN}) we have \setcounter
{CElett}{0}\protect\refstepcounter{equation}\protect\label{eq:St%
ochastic-Final0}\renewcommand\theequation{\arabic{equation}\alph
{CElett}}\setcounter{Ceqindent}{0}\protect\begin{eqnarray}{}^{(2%
)\!}\kappa_0^{\mbox{\scriptsize$N\!$ even}}\:\!\!\hspace{-1.3ex}%
&\displaystyle=&\hspace{-1.3ex}1+\mbox{$\protect\displaystyle
\protect\frac{2}{N^{2\!\!}}$}\,,\protect\stepcounter{CElett}%
\addtocounter{equation}{-1}\protect\label{eq:Stochastic-Ka20Even%
}\setlength{\Cscr}{\value{CEht}\Ctenthex}\addtolength{\Cscr}{-1.%
0ex}\protect\raisebox{0ex}[\value{CEht}\Ctenthex][\Cscr]{}\\*[0e%
x]\protect\displaystyle{}^{(2)\!}\kappa_0^{\mbox{\scriptsize$N\!%
$ odd}}\:\!\!\hspace{-1.3ex}&\displaystyle=&\hspace{-1.3ex}1-%
\mbox{$\protect\displaystyle\protect\frac{1}{N^{2\!\!}}$}\,.%
\protect\stepcounter{CElett}\addtocounter{equation}{-1}\protect
\label{eq:Stochastic-Ka20Odd}\setlength{\Cscr}{\value{CEht}%
\Ctenthex}\addtolength{\Cscr}{-1.0ex}\protect\raisebox{0ex}[%
\value{CEht}\Ctenthex][\Cscr]{}\protect\end{eqnarray}\setcounter
{CEht}{10}\renewcommand\theequation{\arabic{equation}}As noted a%
bove, the expressions (\protect\ref{eq:Stochastic-Final}) are re%
latively ``boring'' functions of $\alpha$. They all approach uni%
ty for \mbox{$\protect\displaystyle\alpha\rightarrow0$}. For 
\mbox{$\protect\displaystyle\alpha\rightarrow\pi/2$}, the expres%
sions (\protect\ref{eq:Stochastic-Ka1Even}) and (\protect\ref{eq%
:Stochastic-Ka2Odd}), involving \mbox{$\protect\displaystyle\cot
\alpha$}, fall smoothly to zero, whereas the expressions (%
\protect\ref{eq:Stochastic-Ka1Odd}) and (\protect\ref{eq:Stochas%
tic-Ka2Even}) rise to the finite values \mbox{$\protect
\displaystyle\pi/2$} and \mbox{$\protect\displaystyle{\pi}{}^{%
\raisebox{-0.25ex}{$\scriptstyle2$}}\!/4$} respectively.\par
\refstepcounter{section}\vspace{1.5\baselineskip}\par{\centering
\bf\thesection. Conclusions\\*[0.5\baselineskip]}\protect\indent
\label{sect:Conclusions}We have found that it is relatively easy 
to get a good physical understanding of the SLAC derivative oper%
ators on \mbox{}\protect\/{\protect\em finite\protect\/} one-dim%
ensional lattices, in position space, by considering explicitly 
the imposition of periodic boundary conditions on the correspond%
ing expressions for infinite lattices. In all cases this leads t%
o small adjustments of coefficients, which can be written down i%
n a very simple form. There are qualitative differences between 
these correction factors depending on whether the lattice has an 
odd or an even number of sites, but in all cases the factors are 
numbers of order unity. This means that the stochastic implement%
ation of the operators proposed in [\ref{cit:Costella2002}]\ req%
uires only superficial changes in order to make the operators co%
mpletely optimal for any finite lattice.\par\vspace{1.5%
\baselineskip}\par{\centering\bf Acknowledgments\\*[0.5%
\baselineskip]}\protect\indent Grateful thanks are due to Robert 
Israel (Department of Mathematics, University of British Columbi%
a) for supplying me with the the general method for evaluating i%
nfinite sums of the form (\protect\ref{eq:Exact-InfiniteSumLinea%
rs1}), (\protect\ref{eq:Exact-InfiniteSumLinears2}), (\protect
\ref{eq:Exact-InfiniteSumSquared1})\ and (\protect\ref{eq:Exact-%
InfiniteSumSquared2}).\par\vspace{1.5\baselineskip}\par{%
\centering\bf References\\*[0.5\baselineskip]}{\protect\mbox{}}%
\vspace{-\baselineskip}\vspace{-2ex}\settowidth\CGDnum{[\ref{cit%
last}]}\setlength{\CGDtext}{\textwidth}\addtolength{\CGDtext}{-%
\CGDnum}\begin{list}{Error!}{\setlength{\labelwidth}{\CGDnum}%
\setlength{\labelsep}{0.75ex}\setlength{\leftmargin}{\labelwidth
}\addtolength{\leftmargin}{\labelsep}\setlength{\rightmargin}{0e%
x}\setlength{\itemsep}{0ex}\setlength{\parsep}{0ex}}\protect
\frenchspacing\setcounter{CBtnc}{1}\addtocounter{CBcit}{1}\item[%
\hfill{[}\arabic{CBcit}{]}]\renewcommand\theCscr{\arabic{CBcit}}%
\protect\refstepcounter{Cscr}\protect\label{cit:Costella2002}J.~%
P.~Costella, \renewcommand\theCscr{Costella}\protect
\refstepcounter{Cscr}\protect\label{au:Costella2002}\renewcommand
\theCscr{2002}\protect\refstepcounter{Cscr}\protect\label{yr:Cos%
tella2002}{}%
\verb+hep-lat/0207008+.\addtocounter{CBcit}{1}\item[\hfill{[}%
\arabic{CBcit}{]}]\renewcommand\theCscr{\arabic{CBcit}}\protect
\refstepcounter{Cscr}\protect\label{cit:Drell1976}S.~D.~Drell, M%
.~Weinstein\ and S.~Yankielowicz, \renewcommand\theCscr{Drell, W%
einstein\ and Yankielowicz}\protect\refstepcounter{Cscr}\protect
\label{au:Drell1976}\renewcommand\theCscr{1976}\protect
\refstepcounter{Cscr}\protect\label{yr:Drell1976}\mbox{}\protect
\/{\protect\em Phys. Rev.~D\protect\/}\ {\bf14}\ (1976) 487; 162%
7.\renewcommand\theCscr{\arabic{CBcit}}\protect\refstepcounter{C%
scr}\protect\label{citlast}\settowidth\Cscr{~[\ref{cit:Drell1976%
}]}\end{list}\par\end{document}